\documentclass[english]{article}
\usepackage{amsmath}
\usepackage{amssymb}
\usepackage{cite}
\usepackage{graphicx}
\usepackage{subfigure}

\newcommand{\e}{\mbox{e}}
\makeatletter

\usepackage{babel}

\begin{document}
\begin{titlepage}
\thispagestyle{empty}

\bigskip

\begin{center}
\noindent{\Large \textbf
{Generalized non-minimal couplings in Randall-Sundrum scenarios}}\\

\vspace{0,5cm}

\noindent{G. Alencar ${}^{a}$\footnote{e-mail: geova@fisica.ufc.br }, I. C. Jardim ${}^{a}$, R. R. Landim ${}^{a}$, C. R. Muniz ${}^{b}$ and R.N. Costa Filho ${}^{a}$}

\vspace{0,5cm}

 {\it ${}^a$Departamento de F\'{\i}sica, Universidade Federal do Cear\'{a}-
Caixa Postal 6030, Campus do Pici, 60455-760, Fortaleza, Cear\'{a}, Brazil. 
 }
 \\
 {\it ${}^b$Universidade Estadual do Cear\'a, Faculdade de Educa\c c\~ao, Ci\^encias e Letras de Iguatu. Iguatu-CE, Brazil.
 }

\end{center}

\vspace{0.3cm}

\begin{abstract}

The Geometrical Localization mechanism in Randall-sundrum (RS) scenarios is extended by considering the coupling between a quadratic mass term and geometrical tensors. Since
the quadratic term is symmetric, tensors with two symmetric indices have to be taken into account. These are the Ricci and the Einstein tensors. For the Ricci tensor it is shown that a localized zero
mode exists while that is not possible for the Einstein tensor.  It is already known that the Ricci scalar generates a localized solution but the metrics do not. Therefore, it can be conclude that divergenceless tensors do not localize the zero mode of gauge
fields. The result is valid for any warp factor recovering the RS metrics at the boundaries, and therefore is valid for RS I and II models. We also compute resonances for all couplings. 
These are calculated using the transfer matrix method. The cases studied consider the standard RS with delta-like branes, and branes generated by kinks and domain-wall as well. The parameters are changed to control the thickness of the smooth brane.
We find that, for all cases considered, geometrical coupling does
not generate resonances. This enforces similar results for the coupling with the Ricci scalar and points to 
the existence of some unidentified fundamental structure of these couplings. 

\end{abstract}
\end{titlepage}

\section{Introduction}

Since the introduction of one compact extra dimension in General Relativity by 
Kaluza and Klein in the $20$'s, the interest in this subject has oscillated during the 
last century \cite{Kaluza:1984ws,Appelquist:1988fh}. The attention to the use of compact extra dimensions changed 
in the late $90$'s when membranes were proposed in the string theory context as 
a contour condition for open strings\cite{Polchinski:1998rq,Polchinski:1998rr}. The main result of these studies is 
the conception of a high dimensional space with extra dimensions where the 
four dimensional world being a surface. In this direction, at the beginning of 
this century, Randall and Sundrum (RS) proposed two models with warped 
geometry in AdS space and delta-like branes \cite{Randall:1999ee,Randall:1999vf}. The so called type I model 
is compact with $Z_2$ symmetry and was posed to solve the hierarchy problem. 
The type II is an alternative to compactification, with a large extra dimension. 
The addition of large extra dimensions brings a new problem: the confinement 
of the fields in the membrane. If the fields are not confined and able to scape 
to the extra dimensions this would be observed in the four dimensional world. 
In these models the confinement of the fields are treated in a very different way 
than in KK compactification, since the fields can not be expanded in discrete 
modes. However, one can make a standard separation of variables, in the five 
dimensional field equations, and obtain an one dimensional Schr\"odinger-like 
equation where the eigenvalues are the mass of the field under consideration. 
Since the free action is quadratic in the fields, the confinement is related to 
the square integrability of the solution to the Schr\"odinger-like equation. The 
zero mode of the gravity field was shown to be localized by RS, providing a 
consistent massless theory of gravity in four dimensions. However, it was shown 
in the original paper that the zero mode of the gauge field is not localized due to 
its conformal invariance. As a consequence this became a drawback of the type 
II model. For non-abelian gauge fields the problem is even worse. For this case 
there are cubic and quartic terms, and if one uses a standard gauge invariant 
theory in the bulk, the gauge invariance will be lost in four dimensions during the 
process of dimensional reduction since for the zero mode  $\int \psi^2 \neq \int \psi^3 \neq \int \psi^4$.

No matter the zero modes are localized or not, resonances can be seen over 
the membrane as possible unstable massive modes. This has been considered in a 
seminal paper  in the same year of the original RS paper \cite{Gremm:1999pj}, and was done by the 
construction of a smooth version of the model to treat the singularity. After this, 
other smooth versions have been considered such as topological defects
and kinks \cite{Bazeia:2005hu,Bazeia:2004yw,Bazeia:2003qt,Liu:2009ve,Zhao:2009ja,Liang:2009zzf,Zhao:2010mk,Zhao:2011hg, Ahmed:2013lea, Guo:2011wr, Dzhunushaliev:2009va, Movahed:2007ps, Du:2013bx,Bazeia:2004dh,Bazeia:2006ef,Csaki:2000fc, German:2012rv} and resonances has been computed. On this subject, a series of papers proposed a  tool, commonly used in condensed matter, to analyze resonances: the
transfer matrix method \cite{Landim:2011ki,Landim:2011ts,Alencar:2012en}. With this tool a rich structure of resonances
has been observed very clearly \cite{Alencar:2010hs,Landim:2010pq}. Despite the complicated form of the
effective potential in smooth versions, analytical solutions has been
found \cite{Cvetic:2008gu}, and it was soon generalized \cite{Landim:2013dja,Landim:2015paa}. With
this solution the Transfer Matrix Method was tested and show to agree
with the analytical results.

Using a different approach, many models has been proposed to solve the problem of zero mode localization
of the $U(1)$ gauge field. Most of them use 
new degrees of freedom, and although consistent, this seems unnatural \cite{Kehagias:2000au,Dvali:1996xe,Chumbes:2011zt}. In the search for models without new
degrees of freedom, Oda proposed that a topologically massive gauge
field could solve the problem of zero mode localization \cite{Oda:2001ux}. However, a
massive gauge field is obtained over the membrane. Following this,
recently Ghoroku {\it et al}. proposed the addition of two terms to the action:
a standard mass term and a boundary mass term \cite{Ghoroku:2001zu}. By tuning the parameters,
a localized zero mode is obtained. Recently, this model was generalized
to include $p-$ form fields and resonances was computed by the present
authors \cite{Jardim:2014vba}. In all this models, although the five dimensional gauge invariance
is lost in five dimension, it is recovered for the zero mode in four
dimensions. But, soon it became clear that this mechanism does not
work for non-abelian gauge fields and Batell {\it et al}. has shown that
a boundary coupling to the field strength is also necessary \cite{Batell:2006dp}. 

Although the Ghoroku {\it et al}. and Batell {\it et al}. models provided the localization of
the zero mode for gauge fields, it was not clear what was the origin
of such couplings. The origin of these couplings were recently proposed \cite{Alencar:2014moa}, where the first was obtained
by coupling the mass term with the Ricci scalar. Both of the Ghoroku {\it et al}. model terms can be naturally generated when delta
like membranes were considered. More than just generating the right couplings, it
was discovered that the model also provided a simple analytical
solution for the zero mode of the Schr\"odinger-like equation. Such
analytical solution has the property of being valid for any conformal
metric, being therefore valid for both, Type I and II RS scenarios. The interesting fact about this solution is that no degree
of freedom was added, providing a more natural extension of the original
RS model. Because of this, the mechanism has been called  "Geometrical
Localization". Interestingly, the same mechanism was proven to work for $p-$forms and Elko spinors \cite{Alencar:2014fga, Jardim:2014cya, Jardim:2014xla}. Moreover, by tuning the value of the couplings,
analytical solutions were also found to these cases and localization
of the zero mode was reached. With this generalization, a consistent vector field can be obtained from the
dimensional reduction of the two-form field in five dimensions. Apparently
the addition of a non-minimal coupling with gravity seems to bring a
bunch of interesting new results to the RS scenario. For example, in a
recent paper the present authors realized a detailed study of resonances
and obtained another very curious numerical result: Geometrical Localization
do not generate resonances for the $p-$forms \cite{Jardim:2015vua}. This is very curious
since this is not the common behavior in theses scenarios. 

When considering non-abelian fields the localization problem is not solved by coupling the mass term with
the Ricci scalar. For completeness of the Geometrical Localization
model, it should be possible to find a geometrical origin of the Batell {\it et al}.
model, which couples the field strength with the boundary. In fact,
the gravity origin of this coupling has been proposed recently in \cite{Germani:2011cv}
based on a non-minimal generalization of the gauge field proposed
some time ago by Horndeski \cite{Horndeski:1976gi}. However, the model does not provide
the localization of the zero mode. To solve this, a recent paper
proposed a model which introduces the Horndeski coupling plus the
coupling with the mass term \cite{Alencar:2015awa}. With this we get a model that reduces
to the Batell {\it et al}. model when a delta like brane is considered. Just as
before, this model also provides analytical solutions, for arbitrary warp factor, to the zero
mode of the Schr\"odinger-like. In this way, the problem
of zero mode localization of $U(N)$ gauge fields is solved, providing
a consistent gauge invariant theory in four dimensions for Type I and II RS models. An important point to be considered is that if the model is valid beyond first approximation. If we consider the full theory the coupling with Ricci scalar will be a function of the brane coordinates and this would spoil gauge invariance. In a recent manuscript \cite{Alencar:2015rtc} we showed that when we consider the full theory in fact gauge invariance is broken throughout an effective mass term. Since the Ricci scalar is coupled to the square of the gauge field, we will have a term proportional to $R(x)A^{2}$ over the brane. Therefore the effective mass is given by $m_\gamma=(\hbar/c)\sqrt{R}/4$. Using Einstein equation we know that $R=8G\pi G\rho/c^2 +4\Lambda$. Where $\rho$ is the density of energy and $\Lambda$ is the cosmological constant. This is very small and far below the present experimental upper bounds of the photon mass, $m_{\gamma}$. For example, for the solar wind magnetic field $B\thicksim10^{-10}$ T, which yields the lower present experimental constraint ($m_{\gamma}\lesssim10^{-55}kg$) \cite{Groom:2000in,Goldhaber:2008xy}, we find that the photon mass is $m_{\gamma}=2\times10^{-72}$ kg, considering the solar magnetosphere as a perfect fluid in the Minkowskyan vacuum. This guarantees the validity of the model. Despite the fact that we have not carefully analyzed the non-abelian case we can estimate the consistence with experimental bounds. In our recently published paper \cite{Alencar:2015rtc}, we have estimated the photon mass in the environment of a neutron star core as being ~ $10^{-47}$ kg. However, we may think that such a mass is more appropriated related to the gluon one, since in that medium prevails this particle as well as quarks (e.g., quark-gluon plasma). We can compare this with the upper bounds for the gluon mass obtained in the literature, which go from $10^{-40}$ to $10{-30}$ kg  \cite{Nussinov:2010jg}. Therefore, the estimation based on our model satisfactorily suits to these limits.

In this manuscript we put forward the effort to understand the
structure of Geometrical Localization. For this we generalize the
previous coupling of the mass term with the Ricci scalar \cite{Alencar:2014moa}. This
is done by considering all the geometrical tensors which can be coupled
with a mass-like term. Since this term is quadratic and symmetric
one must consider tensors with two symmetric indices. The possibilities are the Ricci tensor $R^{MN}$ and the Einstein tensor $G^{MN}$. With this, the localizability of the zero mode of the gauge and of the reduced scalar
fields can be studied. A detailed study of the massive modes by computing
the resonances using the Transfer Matrix method is also performed. 

The paper is organized as follows. The second section presents a short review of the Geometrical
Localization mechanism with the Ricci scalar and how the transversal
and longitudinal components of the gauge fields can be decoupled to
obtain the Schr\"odinger-like equation. In the third section all the needed geometrical objects used in the manuscript will be computed explicitly . The fourth section considers the coupling of the Ricci tensor  with the mass term and solve the mass equation analytically for the zero mode. In the fifth section we study the resonances for this coupling, considering three types of warp factors. In the sixth and seventh sections perform the same steps to the coupling with the Einstein tensor. Finally
in the eighth section we present the conclusion and perspectives.

\section{Review of the Geometrical Localization mechanism}

In this section a short review of the Geometrical Localization
mechanism is presented \cite{Alencar:2014moa}. This mechanism has been achieved by considering
the addition of a coupling between a mass term of a vector field and the Ricci scalar.
Therefore, in five dimensions the gauge invariance is lost. This also breaks the conformal symmetry of the gauge field. However, it has been shown that a massless, gauge invariant, zero mode of the vector field can be obtained over the membrane. Below we
describe how this can be reached. The action is given by

\begin{eqnarray}
 & S_{A}=-\int d^{5}X\sqrt{-g}(\frac{1}{4}g^{MN}g^{PQ}Y_{MP}Y_{NQ}\nonumber \\
 & +\frac{1}{2}\gamma Rg^{MN}X_{M}X_{N})
\end{eqnarray}
where $X_{M}$ is the vector field, $Y_{MN}=\partial_{M}X_{N}-\partial_{M}X_{N}$,
the metric $g^{MN}$ is defined by $ds^{2}=e^{2A(z)}(dx_{\mu}dx^{\mu}+dz^{2})$
and $R$ is the Ricci scalar. The equations of motion are therefore
\begin{equation}
\partial_{M}(\sqrt{-g}g^{MO}g^{NP}Y_{OP})=\gamma \sqrt{-g}Rg^{NP}X_{P}.\label{eqmotion}
\end{equation}

The task now is to show that the above equations will provide a localized
zero mode. For this we must show that the transversal and longitudinal
components of the vector field are decoupled. First of all, from the
antisymmetry of Eq. (\ref{eqmotion}) we obtain the divergenceless
condition $\partial_{N}(\sqrt{-g}RX^{N})=0$. Then split the field
in two parts $X^{\mu}=X_{L}^{\mu}+X_{T}^{\mu}$, where $L$ stands
for longitudinal and $T$ stands for transversal with $X_{T}^{\mu}=(\delta_{\nu}^{\mu}-\frac{\partial^{\mu}\partial_{\nu}}{\Box})X^{\nu}$
and $X_{L}^{\mu}=\frac{\partial^{\mu}\partial_{\nu}}{\Box}X^{\nu}$. With this and $Y_{L}^{5\mu}\equiv X_{L}^{'\mu}-\partial^{\mu}\Phi$
we can show the following identities 
\begin{equation}\label{identities}
\partial_{\mu}Y^{\mu\nu}=\Box X_{T}^{\nu};Y^{5\mu}=X_{T}^{'\mu}+Y_{L}^{5\mu};Y_{L}^{\mu5}=\frac{\partial^{\mu}}{\Box}\partial_{\nu}Y^{\nu5},
\end{equation}
where $\Phi\equiv X_{5}$, the prime means a $z$ derivative and all
lower dimensional indexes will be contracted with $\eta^{\mu\nu}$, and the divergenceless condition can be written as
\begin{equation}
e^{3A}R\partial_{\mu}X_{L}^{\mu}=-(e^{3A}R\Phi)'\label{transverserelation}.
\end{equation}
This is an equation relating the scalar field and the longitudinal
component of the vector field being crucial for the modes decoupling. Now the Eq. (\ref{eqmotion}) can be divided in two. For
$N=5$ 
\begin{equation}
\partial_{\mu}Y^{\mu5}-\gamma e^{2A}R\Phi=0\label{M=00003D5}
\end{equation}
and for $N=\nu$ we get 
\begin{equation}\label{M=00003Dmu}
e^{A}\Box X_{T}^{\nu}+(e^{A}\partial X_{T}^{\nu})'-\gamma e^{3A}RX_{T}^{\nu}+(e^{A}Y_{L}^{5\mu})'-\gamma e^{3A}RX_{L}^{\nu}=0.
\end{equation}

Using now Eqs. (\ref{identities}), (\ref{transverserelation}) and (\ref{M=00003D5}) we get 
\[
(e^{A}Y_{L}^{\mu5})'=\gamma \frac{\partial^{\mu}}{\Box}(e^{3A}R\Phi)'=\gamma e^{3A}RX_{L}^{\nu}
\]
and we finally obtain from Eq. (\ref{M=00003Dmu}) the equation for the
transverse part of the gauge field 
\[
e^{A}\Box X_{T}^{\nu}+(e^{A}\partial X_{T}^{\nu})'+\gamma e^{3A}RX_{T}^{\nu}=0.
\]

Now by separating the $z$ dependence like $X_{T}^{\mu}=\tilde{X}_{T}^{\mu}\tilde{\psi}(z)$,
using $R=-4(2A''+3A'^{2})e^{-2A}$ and performing the transformation
$\tilde{\psi}=e^{-\frac{A}{2}}\psi$ we get the desired Schr\"odinger
equation with the potential 
\begin{equation}
U=(\frac{1}{4}-12\gamma)A'^{2}+(\frac{1}{2}-8\gamma)A''\label{effectivepotential},
\end{equation}
which is localized for $\gamma=1/16$ with solution $e^{A}$.
For the scalar field we must be careful since we have 
\[
\Box\Phi-(\partial_{\mu}A^{\mu})'-\gamma Re^{2A}\Phi=0.
\]

Performing the separation of variables $\Phi=\Psi(z)\phi(x)$, defining
$\Psi=(e^{3A}R)^{-1/2}\psi$, and using Eq. (\ref{transverserelation})
we get a Schr\"odinger equation for the
massive mode of the scalar field with a potential given by \cite{Alencar:2014moa}
\[
U=\frac{1}{4}(3A'+(\ln R)')^{2}-\frac{1}{2}(3A''+(\ln R)'')+\gamma Re^{2A}.
\]
With this potential we see that the zero mode of the scalar field
solution is localized for $\gamma=9/16$. This shows us that we
cannot have both fields localized.

When considering the reduced scalar field an equation will appear throughout the manuscript, therefore lets manipulate it in the general form
\begin{equation}\label{scalargeneral}
\partial(e^{(-f-cA)}\partial(e^{(g+cA)}\Psi))+h\Psi=-m^{2}\Psi,
\end{equation}
or
\[
-\Psi''-(2g-f+cA)'\Psi'-((g+ca)'e^{(g-f)})'+h)\Psi=m^{2}e^{-(g-f)}\Psi^{2}.
\]

If we compare with the identity found in \cite{Landim:2011ki}
we get $P=-(2g-f+ca))$, $V=-((g+ca)'e^{(g-f)})'-h$ and $Q=e^{-(g-f)}$.
With this we obtain
\[
\frac{dz}{dy}=\Theta(y),\quad\psi(y)=\Omega(y)\overline{\psi}(z),
\]
with
\[
\Theta(y)=e^{-\frac{1}{2}(g-f)},\quad\Omega(y)=e^{\frac{1}{4}f-\frac{3}{4}g-\frac{1}{2}cA}.
\]
and the potential
\[
{U}(z)=V(y)/\Theta^{2}+\left(P'(y)\Omega'(y)-\Omega''(y)\right)/\Omega\Theta^{2}.
\]
With our definition we see that
\[
\Omega'=-(\frac{3}{4}g-\frac{1}{4}f+\frac{1}{2}cA)'\Omega;\Omega''=e^{\frac{3}{4}g-\frac{1}{4}f+\frac{1}{2}cA}((-f-cA)^{-\frac{1}{4}}(g+ca)^{-\frac{3}{4}})''\Omega,
\]
and the final form of the potential is
\[
U(z)=-\frac{1}{2}(cA+\frac{1}{2}(f+g))''+\frac{1}{4}(cA+f)'^{2}-\frac{1}{16}(g-f)'^{2}.
\]

Before considering the generalization to other couplings, in the next section we must yet review and obtain the necessary geometrical objects to be used in the manuscript.

\section{Geometrical Tensors in RS Scenario}

Due to the variety of geometrical objects needed in this manuscript
we use this section to obtain them. The metric is given by $ds^{2}=e^{2A(z)}\eta_{MN}dx^{M}dx^{N}$ from where 
\[
\Gamma_{55}^{5}=A',\Gamma_{\mu\nu}^{5}=-A'\eta_{\mu\nu,}\Gamma_{5\alpha}^{\mu}=\delta_{\alpha}^{\mu}A',
\]
and the components of the curvature tensor are
\begin{equation}\label{RMNPQgeral}
R_{\mu\nu\alpha5}=0,R_{\mu5\nu5}=-\eta_{\mu\nu}A''e^{2A},R_{\mu\nu\alpha\beta}=-A'^{2}e^{2A}(\eta_{\mu\alpha}\eta_{\nu\beta}-\eta_{\nu\alpha}\eta_{\mu\beta}),
\end{equation}
leading to the Ricci tensor and scalar
\begin{equation}\label{RMNRgeral}
R_{\mu\nu}=-\eta_{\mu\nu}(A''+3A'^{2}),R_{55}=-4A'',R=-e^{-2A}(8A''+12A'^{2}).
\end{equation}

The other tensor that will be used in this manuscript is the Einstein tensor that now can be  easily obtained 
\begin{equation}\label{GMNgeneral}
G_{55}=6A'^{2};G_{\mu\nu}=3\eta_{\mu\nu}(A''+A'^{2}).
\end{equation}

In subsequent sections we will consider warp factors which asymptotically
recovers the RS metric, namely $A(z)=-\ln (k|z|+1)$. Therefore, we also give here explicitly all the geometrical
quantities for this case. First we should note that
\[
A'=-k\theta(z)e^{A};A''=k^{2}e^{2A}-2k\delta(z)e^{A},
\]
and we get, by using the fact that $f(x)\delta(x)=f(0)\delta(x)$,
\[
R_{\mu\nu\alpha5}=0,R_{\mu5\nu5}=-\eta_{\mu\nu}e^{4A}(k^{2}-2k\delta(z)),R_{\mu\nu\alpha\beta}=k^{2}e^{4A}(\eta_{\nu\alpha}\eta_{\mu\beta}-\eta_{\nu\beta}\eta_{\mu\alpha}).
\]
From the above we get
\[
R_{\mu\nu}=-\eta_{\mu\nu}e^{2A}(-2k\delta(z)+4k^{2}),R_{55}=-(4k^{2}-8k\delta(z))e^{2A},
\]
and
\[
R=16k\delta(z)-20k^{2}.
\]
 
Due to the simplicity of the expressions (\ref{RMNRgeral}) and (\ref{GMNgeneral}) we can yet simplify
it relating to trace expressions by
\begin{equation}\label{traceRMN}
R_{\mu\nu}=\frac{\eta_{\mu\nu}}{4}e^{2A}\bar{R}_{\;\alpha}^{\alpha},R_{55}=e^{2A}\bar{R}_{\;5}^{5},
\end{equation}
and 
\begin{equation}\label{traceGMN}
G_{\mu\nu}=\frac{\eta_{\mu\nu}}{4}e^{2A}\bar{G}_{\;\alpha}^{\alpha},G_{55}=e^{2A}\bar{G}_{\;5}^{5}.
\end{equation}

Where we have used a bar to indicate that an index has been raised with $g^{\mu\nu}$, namely
\begin{equation}
\bar{R}_{\quad\alpha}^{\alpha}=g^{\mu\nu}R_{\mu\nu};\bar{R}_{\quad5}^{5}=g^{55}R_{55},
\end{equation}
and
\begin{equation}
\bar{G}_{\quad\alpha}^{\alpha}=g^{\mu\nu}G_{\mu\nu};\bar{G}_{\quad5}^{5}=g^{55}G_{55}.
\end{equation}

The relation between objects with and without bar is given by $\bar{T}_{\quad\alpha}^{\alpha}=e^{-2A}T_{\quad\alpha}^{\alpha}$. In the next section we must use the above results to study a variety
of geometrical couplings giving localized modes for the
fields.

\section{The Coupling to the Ricci Tensor $R^{MN}$}

In this section we show that 
the longitudinal and transversal parts of the gauge field decouples, and we also analyze the localizability 
of the zero mode. The study of the massive modes will be left to the next section. 
The action considered in this case is given by 
\footnote{Here the signal used is different than the one in\cite{Alencar:2014fga} for the mass
term. That does not change the result.}
\begin{equation}
S_{1}=-\int d^{5}X\sqrt{-g}(\frac{1}{4}g^{MN}g^{PQ}Y_{MP}Y_{NQ} -\frac{\gamma_{1}}{2}\int d^{5}x\sqrt{-g}R^{MN}X_{M}X_{N}.
\end{equation}
In the above action $\gamma_{1}$ is the coupling constant. In this and in the next section we are going to use subscripts in order to avoid confusion. With this we get the equations of motion
\begin{equation}\label{EOMRMN}
\partial_{M}(\sqrt{-g}g^{MO}g^{NP}Y_{OP})=\gamma_{1}\sqrt{-g}g^{NO}g^{MP}R_{OM}X_{P},
\end{equation}
and from the antisymmetry of $Y$ we get the divergence condition
\begin{equation}\label{divergenceRMN}
\partial_{N}(\sqrt{-g}g^{NO}g^{MP}R_{OM}X_{P})=0.
\end{equation}
Since the gauge freedom is lost, one must be careful in dealing with
the above equations. First, for $N=\nu$ and using Eq. (\ref{EOMRMN}) we get 
\[
e^{A}\partial_{\mu}Y^{\mu\nu}+\partial(e^{A}Y^{5\nu})=\gamma_{1}e^{A}R_{\;\mu}^{\nu}X^{\mu},
\]
where for $N=5$ we obtain
\[
\partial_{\mu}(Y^{\mu5})=\gamma_{1}R_{\;5}^{5}\Phi.
\]

In the above equations we have used our previous definition $A^5=\Phi$. From now on all the indices will be raised
with $\eta^{\mu\nu}$. The divergence condition becomes 
\[
\partial_{\nu}(e^{A}R_{\;\mu}^{\nu}X^{\mu})=-\partial(e^{A}R_{\;5}^{5}\Phi).
\]
We have to point that, just as in Ref. \cite{Alencar:2014moa}, we must show that the
longitudinal part decouples from the transversal one as defined in Section two. Using the 
identities (\ref{identities}) we see that our equations of motion become
\begin{equation}\label{M5L}
\partial_{\mu}Y_{L}^{\mu5}-\gamma_{1}R_{\;5}^{5}\Phi=0,
\end{equation}
and
\[
e^{A}\Box X_{T}^{\nu}+\partial\left(e^{A}\partial X_{T}^{\nu}\right)-\gamma_{1}e^{A}R_{\;\mu}^{\nu}X_{T}^{\mu}+\partial(e^{A}Y_{L}^{5\mu})-\gamma_{1}e^{A}R_{\;\mu}^{\nu}X_{L}^{\mu}=0.
\]

Now using the last identity of (\ref{identities}) and Eq. (\ref{M5L}) we get the identities
\[
Y_{L}^{\mu5}=\frac{\partial^{\mu}}{\Box}\partial_{\nu}Y^{\nu5}=\gamma_{1}R_{\;5}^{5}\frac{\partial^{\mu}}{\Box}\Phi
\]
and 
\[
\partial(e^{A}Y_{L}^{5\mu})=-\partial(e^{A}Y_{L}^{\mu5})=-\gamma_{1}\frac{\partial^{\mu}}{\Box}\partial(e^{A}R_{\;5}^{5}\Phi)=\gamma_{1}e^{A}\frac{\partial^{\mu}}{\Box}\partial_{\nu}(R_{\;\alpha}^{\nu}X^{\alpha}).
\]

At this point we see that it is not trivial to obtain the desired
result. The only way for obtaining it is if the condition
\begin{equation}\label{conditionRMN}
\partial^{\mu}\partial_{\nu}(R_{\;\alpha}^{\nu}X^{\alpha})=R_{\;\alpha}^{\mu}\partial^{\alpha}\partial_{\nu}X^{\nu}
\end{equation}
is satisfied. This is true for this case since from Eq. (\ref{traceRMN}) we have 
\[
R_{\;\alpha}^{\nu}=R_{\;\beta}^{\beta}\frac{\delta_{\alpha}^{\nu}}{4},
\]
and it can be verified easily that the condition (\ref{conditionRMN}) is in fact valid. With this results we then get the identity
\[
\partial(e^{A}Y_{L}^{5\mu})=\gamma_{1}e^{A}R_{\;\alpha}^{\mu}\frac{\partial^{\alpha}}{\Box}(\partial_{\nu}X^{\nu})=\gamma_{1}e^{A}R_{\;\alpha}^{\mu}X_{L}^{\alpha}=\gamma_{1}e^{A}R_{\;\alpha}^{\mu}X_{L}^{\alpha},
\]
and therefore we obtain for the transverse gauge field the equation
after using again (\ref{traceRMN})
\[
e^{A}\Box X_{T}^{\nu}+\partial\left(e^{A}\partial X_{T}^{\nu}\right)-\gamma_{1}e^{A}\frac{R_{\;\beta}^{\beta}}{4}X_{T}^{\nu}=0.
\]

The mass equation is obtained in the standard way by separating $X_{T}^{\nu}=\tilde{X}_{T}^{\nu}(x^{\mu})\Psi(z)$
and is given by

\begin{equation}\label{eqfp}
 e^{-A}\partial\left(e^{A}\partial\Psi\right)-\gamma_{1}\frac{R_{\;\beta}^{\beta}}{4}\Psi=-m^{2}\Psi
\end{equation}
and the Schr\"odinger equation is obtained by the transformation $\Psi=e^{-\frac{A}{2}}\psi$
with potential
\begin{equation}\label{potpR} 
U=\frac{1}{4}A'^{2}+\frac{1}{2}A''+\gamma_{1}\frac{R_{\;\beta}^{\beta}}{4}
=(\frac{1}{4}-3\gamma_{1})A'^{2}+(\frac{1}{2}-\gamma_{1})A''.
\end{equation}
 Using the ansatz $\psi=e^{aA}$, we can see that, for $\gamma_{1}=-2$, we have
a localized mode zero solution given by $e^{\frac{5}{2}A}$. 

For the scalar
field we have
\begin{equation}\label{scalarphi}
 \Box\Phi-\partial\partial_{\mu}X^{\mu}-\gamma_{1}R_{\;5}^{5}\Phi=0.
\end{equation}
We can see that the above equation is also not trivially decoupled
from $X^{\mu}$ component since the divergence equation now do not
involves the term $\partial_{\mu}X^{\mu}$. However by using identity
(\ref{traceRMN}) we see that 
\[
\partial_{\nu}(R_{\;\alpha}^{\nu}X^{\alpha})=\frac{R_{\;\beta}^{\beta}}{4}\partial_{\mu}X^{\mu}
\]
and the divergenceless condition becomes
\[
\frac{R_{\;\beta}^{\beta}}{4}\partial_{\mu}X^{\mu}=-e^{-A}\partial(e^{A}R_{\;5}^{5}X^{5}).
\]

Now by separating $\Phi=\tilde{\Phi}\tilde{\phi}(z)$ in (\ref{scalarphi}) we get
\[
4\partial[\frac{e^{-3A}}{\bar{R}_{\;\beta}^{\beta}{}_{\;}}\partial(e^{3A}\bar{R}_{5}^{5}\tilde{\phi})]-\gamma_{1}e^{2A}\bar{R}_{5}^{5}\tilde{\phi}=-m^{2}\tilde{\phi}
\]
and comparing with our expression (\ref{scalargeneral}) with $c=3$, $f=\ln\bar{R}_{\;\beta}^{\beta}$
and $g=\ln\bar{R}_{5}^{5}$ we obtain
\begin{equation}\label{potp-1R}
 U(z)=\frac{1}{4}(3A+\ln\bar{R}_{\;\beta}^{\beta})'^{2}-\frac{1}{2}(3A+\frac{1}{2}(\ln\bar{R}_{\;\beta}^{\beta}+\ln\bar{R}_{5}^{5}))''-\frac{1}{16}(\ln\bar{R}_{5}^{5}-\ln\bar{R}_{\;\beta}^{\beta})'^{2}+\gamma_{1}e^{2A}\bar{R}_{5}^{5}.
\end{equation}

Despite the fact that we have not been able to find an analytical
solution to the zero mode of the above potential. We still can obtain some results.
An interesting fact about this potential is that in the limit $z\to\infty$
we get that $\ln\bar{R}_{\;\beta}^{\beta}$ and $\ln\bar{R}_{5}^{5}$
are constants and the potential reduces to the asymptotic form
\[
U(z)=\frac{9}{4}A'^{2}-(\frac{3}{2}+4\gamma_{1})A''.
\]
Therefore, we see that this term is identical to the case with the
$R$ coupling, but the mass term is different. The localized solution
for the above equation is obtained with $\gamma_{1}=-\frac{3}{4}$
and is given by $e^{\frac{3}{2}A}$. It is clear that we can
not localize both fields with this coupling. This enforces our previous
results  that we just can localize one of the fields\cite{Alencar:2014moa,Alencar:2014fga}. 

\section{Resonances for the Ricci Tensor Coupling}
In this section we compute the transmission coefficient for massive modes in some brane scenarios. The presence of resonances peaks in the transmission coefficient indicates the existence of unstable massive modes. 
Except for the RS model, which can be solved analytically, all others were made using the transfer matrix method.

\subsection{The Randall-Sundrum scenario}
The first brane scenario that we will study is the Randall-Sundrum scenario. Despite the singularity this scenario has a historical importance and serves as an important paradigm in physics of extra dimensions and field localization.
The warp factor of this scenario in a conformal form is given by
\begin{equation}
A(z) = -\ln\left[k|z| +1\right].
\end{equation}
In this case the components of Ricci tensor are
\begin{eqnarray}
&& R^{\beta}_{\beta} = -16k^{2} +8k\delta(z),
\\&& R^{5}_{5} =  -4k^{2} +8k\delta(z).
\end{eqnarray}

\begin{figure}[h!]
 \centering
 \includegraphics[scale=0.4]{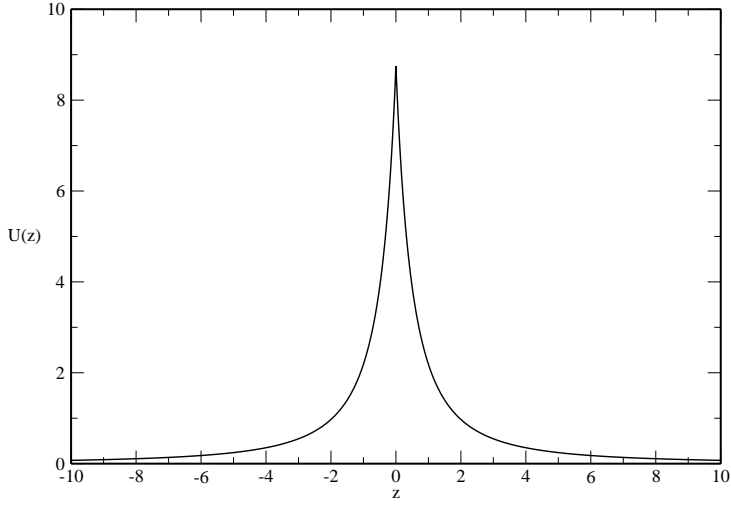}
 \caption{Plot of Schr\"odinger potential for gauge field in Randall-Sundrum scenario with $k =1$.}
 \label{fig:potRmn-RS}
\end{figure}

\begin{figure}[h!]
 \centering
 \includegraphics[scale=0.4]{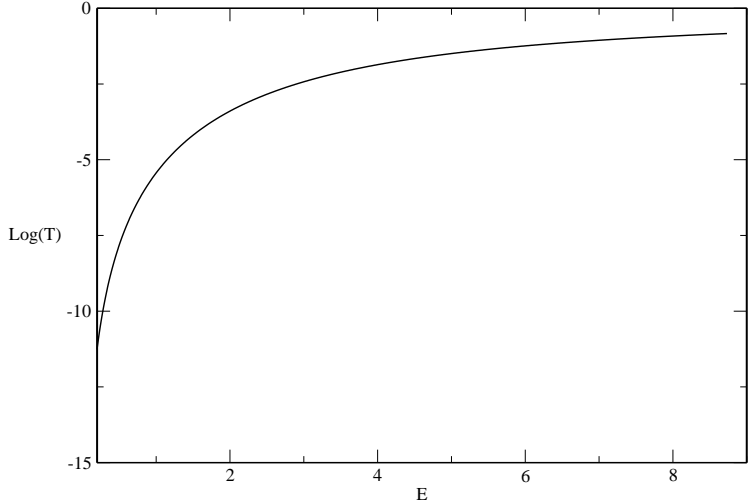}
 \caption{Transmission coefficient for gauge field in Randall-Sundrum scenario with $k=1$ as function of energy, $E = m_{X}^{2}$.}
 \label{fig:TRmn-RS}
\end{figure}

\begin{figure}[!h]
\centering
\includegraphics[scale=0.4]{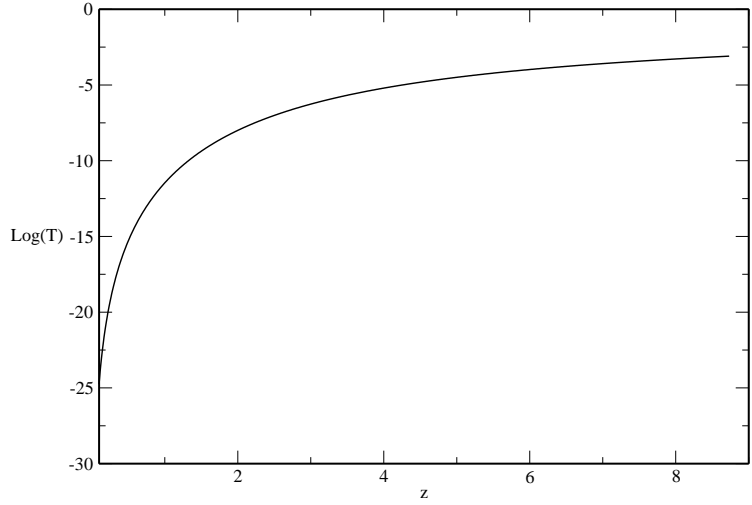}
\caption{Transmission coefficient for scalar field in Randall-Sundrum scenario with $k=1$ as function of energy, $E = m_{0}^{2}$.}
\label{fig:TRmn1-RS}
\end{figure}

For the transversal part of $p$-form the potential of Schr\"odinger equation, Eq. (\ref{potpR}), is given by 
\begin{equation}\label{potRRS}
U(z)=\frac{35k^{2}}{4(k|z|+1)^{2}} -5k\delta(z),
\end{equation}
and was illustrated in fig. \ref{fig:potRmn-RS} for some $p$-forms.
For the massive case  eq. (\ref{eqfp}) provides the solution
\begin{equation}\label{psimasRRS}
\psi(z) =(k|z|+1)^{1/2}[C_{1}J_{\nu}(m_{X}|z|+ m_{X}/k)+C_{2}Y_{\nu}(m_{X}|z|+ m_{X}/k)],
\end{equation}
where $C_{1}$ and $C_{2}$ are constants and $\nu = (2p+1)/2$ . Since the Bessel functions goes to infinity as $(m_{X}|z|+ m_{X}/k)^{-1/2}$, no fixation of constants $C_{1}$ and $C_{2}$ produces a convergent solution. Then the massive modes are non-localized. To obtain more information about massive modes we evaluate the transmission coefficient. For this we will write the solution (\ref{psimasRRS}) in the form
\begin{equation}
 \psi(z) = \left\lbrace\begin{matrix}E_{\nu}(-z)+\sigma F_{\nu}(-z) &,\;\mbox{for}\; z<0 \\ 
\gamma F_{\nu}(z)&,\;\mbox{for}\; z\geq0\end{matrix}\right.,
\end{equation}
where
\begin{eqnarray}
&&E_{\nu}(z) = \sqrt{\frac{\pi}{2}}(m_{X}z+ m_{X}/k)^{1/2}H^{(2)}_{\nu}(m_{X}z+\ m_{X}/k)
\\&& F_{\nu}(z) = \sqrt{\frac{\pi}{2}}(m_{X}z+m_{X}/k)^{1/2}H^{(1)}_{\nu}(m_{X}z+ m_{X}/k),
\end{eqnarray}
and $H^{(1)}_{\nu}$ and $H^{(2)}_{\nu}$ are the Hankel functions of first and second kind respectively. The boundary conditions at $z = 0$ imposes 
\begin{equation}
 \gamma =  \frac{W(E_{\nu},F_{\nu})(0)}{ 2F_{\nu}(0)F_{\nu}'(0) + 2pk F_{\nu}^{2}(0)},
\end{equation}
where $ W(E_{\nu},F_{\nu})(0) =  E_{\nu}(0)F_{\nu}'(0)-E_{\nu}'(0)F_{\nu}(0)$ is the Wronskian  at $z=0$. Since it is constant in Schr\"odinger equation the transmission coefficient can be written as
\begin{equation}
T = |\gamma|^{2} = \frac{m_{X}^{2}}{|F_{\nu}(0)F_{\nu}'(0) + pk  F_{\nu}^{2}(0)|^{2}}. 
\end{equation}
The transmission coefficient was plotted in fig. \ref{fig:TRmn-RS} as function of energy for some $p$-forms and do not show peaks, indicating no unstable massive modes.  
\\
For the scalar field  in Randall-Sundrum scenario the potential of Schr\"odinger equation, (\ref{potp-1R}), can be written as
\begin{equation}\label{potp-1RRS}
U(z)= \frac{35k^{2}}{4(k|z|+1)^{2}} -13k\delta(z).
\end{equation}
Since $R_{reg}$ in RS scenario is a constant, it does not contribute to the potential.
The potential for $(p-1)$-form is the same of $p$-form, changing only the boundary condition, the behavior of massive  modes are the same, i.e, non-localized.
The transmission coefficient was illustrated in fig. \ref{fig:TRmn1-RS}  shows the same behavior of the gauge field.

\subsection{The brane scenario generated a domain-wall}
In order to solve the singularity in RS models, some models with smooth branes appears. Now we compute the transmission coefficient for a smooth brane produced by a domain-wall \cite{Du:2013bx, Melfo:2002wd}.
The warp factor,
\begin{equation}
  A(z) = -\frac{1}{2n}\ln\left[\left(kz\right)^{2n}+1\right],
\end{equation}
recovers the Randall-Sundrum metric for large $z$ and $n \in N^{*}$. This brane scenario produces the following components of Ricci tensor
\begin{eqnarray}
&&R^{\beta}_{\beta} =  -4k^{2} (kz)^{2 n-2} \left((kz)^{2 n}+1\right)^{\frac{1}{n}-2} \left(4 (kz)^{2 n}-2 n+1\right) \label{Rbdw}
\\&&R^{5}_{5} = -4k^{2} (kz)^{2 n-2} \left((kz)^{2 n}+1\right)^{\frac{1}{n}-2} \left((kz)^{2 n}-2 n+1\right). \label{R5dw}
\end{eqnarray}

\begin{figure}[!h]
\centering
\subfigure[]{
\includegraphics[scale=0.2]{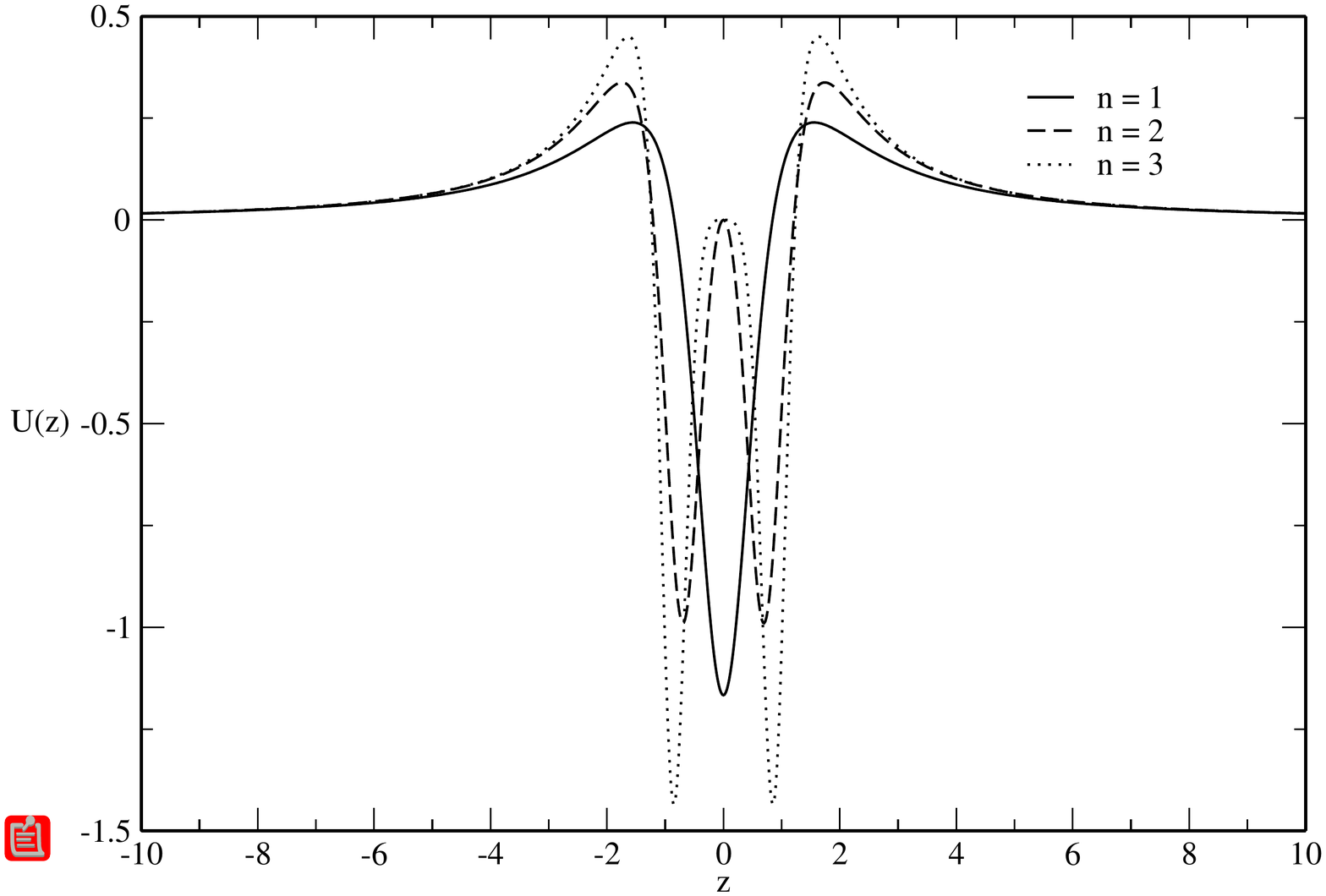}
\label{fig:potRmn-sm}
}
\subfigure[]{
\includegraphics[scale=0.2]{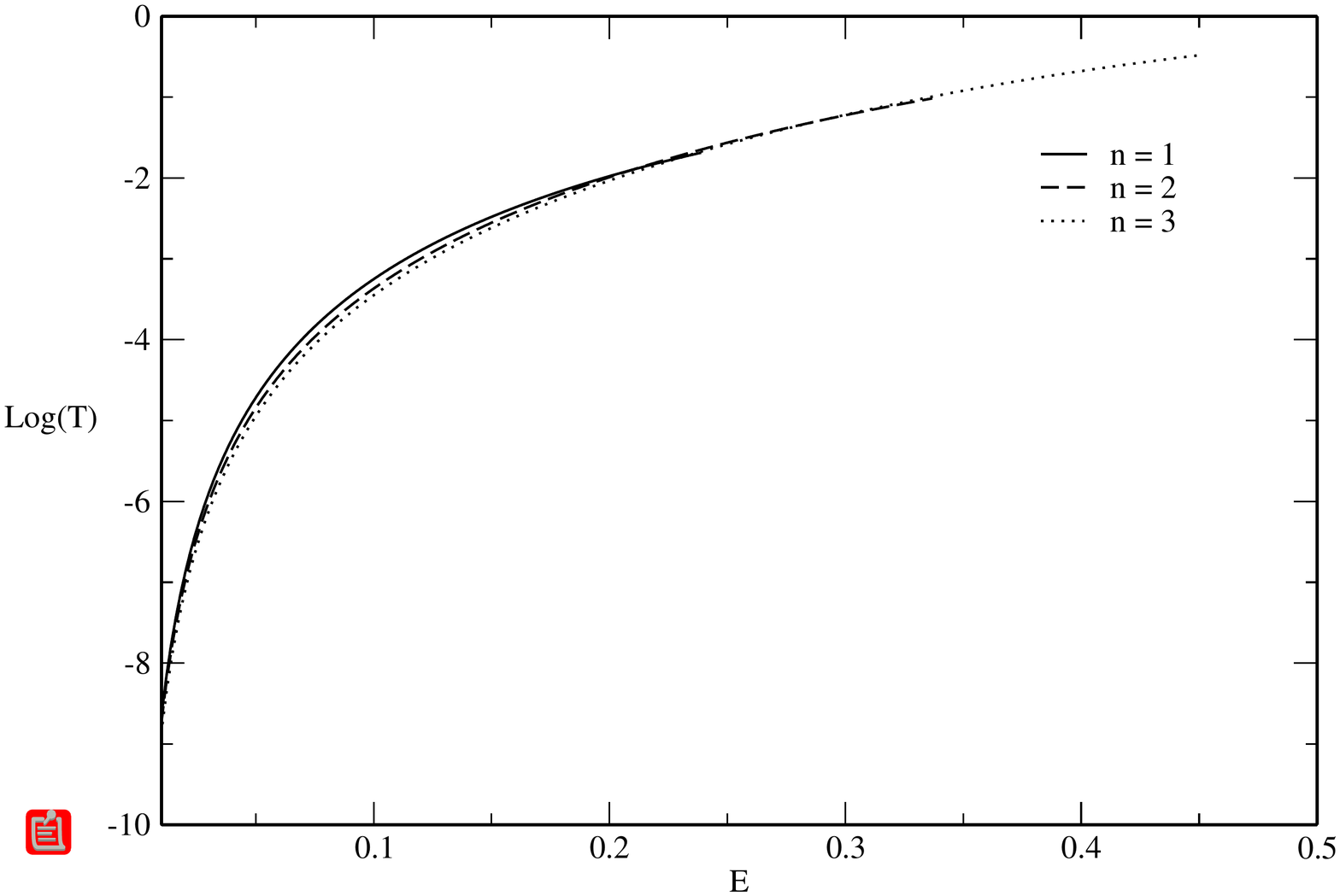}
\label{fig:TRmn-sm}
}
\caption{(a) Schr\"odinger potential in smooth scenario generated by domain walls for some values of parameter $n$. (b)Transmission coefficient in smooth scenario generated by domain walls  for some values of parameter $n$ as function of energy, $E= m_{X}^{2}$.}
\end{figure}

Using this metric in Eq. (\ref{potpR}) we obtain the Schr\"odinger's potential for transversal part of $p$-form
\begin{equation}\label{potRsm}
U(z)=\frac{5k^{2} (k z)^{2 n-2} \left(7 (k z)^{2 n}-4 n+2\right)}{4\left((k z)^{2 n}+1\right)^2},
\end{equation}
which was illustrated in fig. \ref{fig:potRmn-sm} for some values of $n$.

The massive modes solutions for the  transversal gauge field can not be found analytically. To obtain information about this state we use the transfer matrix method to evaluate the transmission coefficient. The behavior is illustrated in fig. \ref{fig:TRmn-sm} for some values of parameter $n$, where there are no peaks indicating the absence of unstable modes.

For the reduced scalar field the potential is given by eq. (\ref{potp-1R}). Because the components of the Ricci tensor vanishes vanishes at regular points near to the origin for all values of parameter
$n$, the potential diverges at these same points. These kind of divergence does not allow us to use the transfer matrix method to compute the transmission coefficient, and to evaluate the existence of 
unstable massive modes.
\subsection{The brane scenario generated a kink} 
\begin{figure}[!h]
\centering
\subfigure[]{
\includegraphics[scale=0.2]{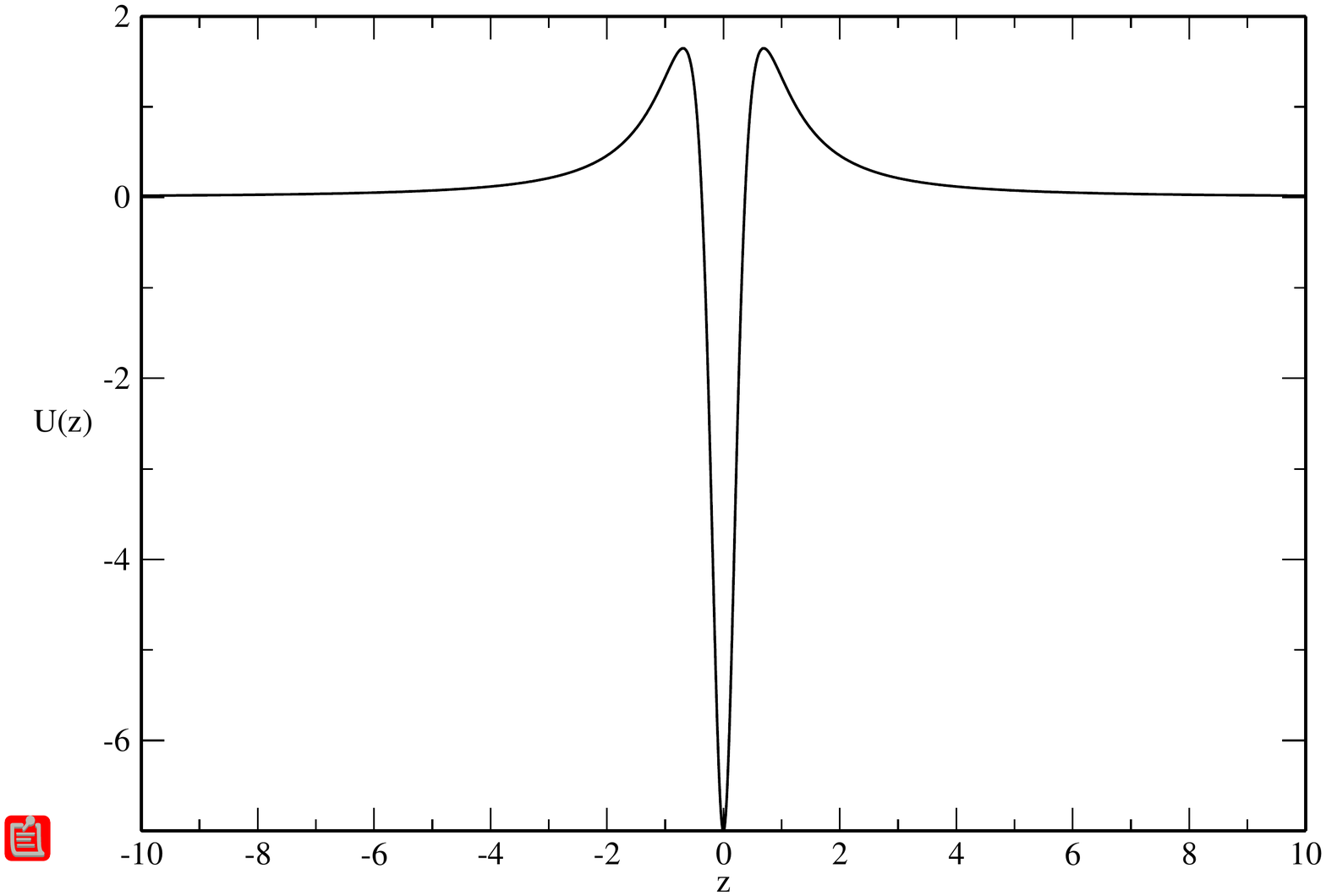}
\label{fig:potRmn-kink}
}
\subfigure[]{
\includegraphics[scale=0.2]{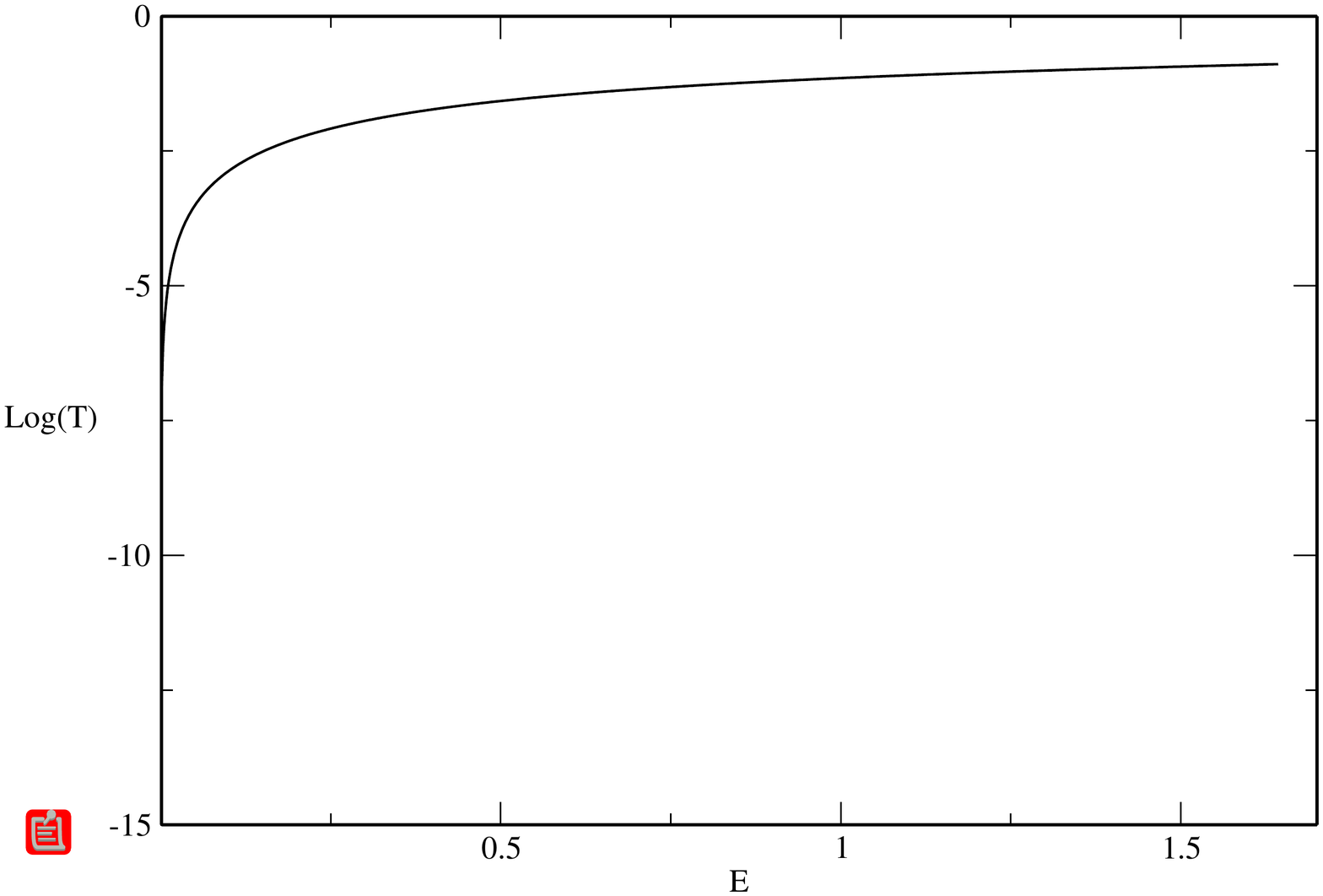}
\label{fig:TRmn-kink}
}
\caption{(a) Schr\"odinger potential in kink scenario for gauge field. (b) Transmission coefficient in kink scenario for gauge field as function of energy, $E= m_{X}^{2}$.}
\end{figure}

Another smooth brane scenario is generated by a kink \cite{Landim:2011ki}. The warp factor is given by 
\begin{equation}
 A(y) = -4\ln\cosh y -\tanh^{2}y,
\end{equation}
where the variable $y$ is related to the conformal coordinate, $z$, by
\begin{equation}
dz = \e^ {-A(y)}dy.
\end{equation}
The potential's behavior of the gauge field transversal part, eq. (\ref{potpR}), with the above warp factor is illustrated in fig. \ref{fig:potRmn-kink}. For massive modes, like in previous sections, we use
the transfer matrix  method to compute the transmission coefficient. The result is plotted in fig. \ref{fig:TRmn-kink}. The figure does no exhibit resonance peaks. 
\\For the reduced scalar field the potential is given by eq. (\ref{potp-1R}). Like the previous case the components of the Ricci tensor vanishes at regular points near to the origin,  the potential diverges at these same points. These kind of divergence does not allow us to use the transfer matrix method to compute the transmission coefficient and to evaluate the existence of unstable massive modes.

\section{The Coupling to the Einstein tensor $G^{MN}$}
In this section we try to generalize the results obtained in the previous section to more general tensors. The
covariant possibilities are the Ricci tensor $R^{MN}$, the curvature
tensor $R^{MNPQ}$ and the Einstein tensor $G^{MN}$. The curvature
tensor would need a fourth order interaction term generating a nonlinear
equation and we do not consider it.  Therefore, we are left with the possibility of the Einstein tensor $G^{MN}$. This tensor is a combination of the cases considered previously. 
Since $G^{MN}$ is the only possible combination of $g^{MN}R$ and $R^{MN}$ with null divergence. We can analyze our results for two kinds of tensors: $g^{ MN} $ and $G^{MN}$ with null divergence,  and $g^{MN}R$ and $R^{MN}$ with non null divergence. The action
is given by

\begin{eqnarray}
 & S_{1}=-\int d^{5}X\sqrt{-g}(\frac{1}{4}g^{MN}g^{PQ}Y_{MP}Y_{NQ}\nonumber \\
 & -\frac{\gamma_{2}}{2}\int d^{5}x\sqrt{-g}G^{MN}X_{M}X_{N}.
\end{eqnarray}
The Einstein tensor is just a combination of the Ricci tensor and Ricci scalar. Using also the fact that $G_{\mu\nu}$ satisfies a decoupling condition
similar to (\ref{conditionRMN})
\[
\partial_{\mu}(G^{\mu\nu}X_{\nu})=\frac{G_{\mu}^{\mu}}{4}\partial_{\mu}X^{\mu},
\]
we get that the longitudinal and transversal parts that can be easily decoupled by following the same steps as before. Therefore,
we just have to sum the effective potentials to get
\begin{equation}\label{potpG}
 U=(\frac{1}{4}+3\gamma_{2})A'^{2}+(\frac{1}{2}+3\gamma_{2})A''.
\end{equation}

The condition for localization is $\gamma_{2}=0$ and therefore
we get a non-localized zero mode. Now we must consider the scalar
field. Again the case is very similar to the one of the last section. We
just need to change $\bar{R}_{\;\alpha}^{\alpha}\to\bar{G}_{\;\alpha}^{\alpha}$ and 
$\bar{R}_{\;5}^{5}\to\bar{G}_{\;5}^{5}$ to get the final effective potential
\begin{equation}\label{potp-1G}
 U(z)=-\frac{1}{2}(3A+\frac{1}{2}(\ln\bar{G}_{\;\beta}^{\beta}+\ln\bar{G}_{5}^{5}))''+\frac{1}{4}(3A+\ln\bar{G}_{\;\beta}^{\beta})'^{2}-\frac{1}{16}(\ln\bar{G}_{5}^{5}-\ln\bar{G}_{\;\beta}^{\beta})'^{2}+\gamma_{2}e^{2A}\bar{G}_{5}^{5}.
\end{equation}

Just as in the last cases we did not find an analytical solution.
However we can study localizability by studying its asymptotic behavior.
Again we have that for large $z$, $\bar{G}_{\;\beta}^{\beta}$ and
\textbf{$\bar{G}_{5}^{5}$} are constants and we get

\[
U(z)=(\frac{9}{4}+6\gamma_{2})A'^{2}-\frac{3}{2}A''.
\]
We get $\gamma_{2}=0$ as a solution and we can not localize the
scalar field. Despite the fact that the zero mode is not localized, in the next section we must consider the resonances for this kind of coupling.

\section{Resonances for the Einstein Tensor Coupling}
Since the zero mode of gauge field coupled with Einstein tensor is non-localized there is no fixation of the coupling constant. Therefore the coupling with Einstein tensor has a free parameter. Despite the fact that the zero mode is not localized we can study the dependence of the resonances of the massive modes on this parameter. The curious fact is that this coupling modifies the model in such a way that no resonances is found for any value of $\gamma_2$. Unlike the previous case, the coupling constant must to be positive to provide a positive maximum for the potential and a positive asymptotic behavior. 
\subsection{The Randall-Sundrum scenario}
As in the case of coupling with Ricci tensor, we first compute the transmission coefficient in RS scenario, because this is the only case that can be solved analytically. 
The warp factor of this scenario in a conformal form is given by
\begin{equation}
A(z) = -\ln\left[k|z| +1\right].
\end{equation}
In this case the components of Einstein tensor are
\begin{eqnarray}
&& G^{\beta}_{\beta} = 24k^{2} -24k\delta(z),
\\&& G^{5}_{5} =  6k^{2} -12k\delta(z).
\end{eqnarray}

\begin{figure}[h!]
 \centering
 \includegraphics[scale=0.4]{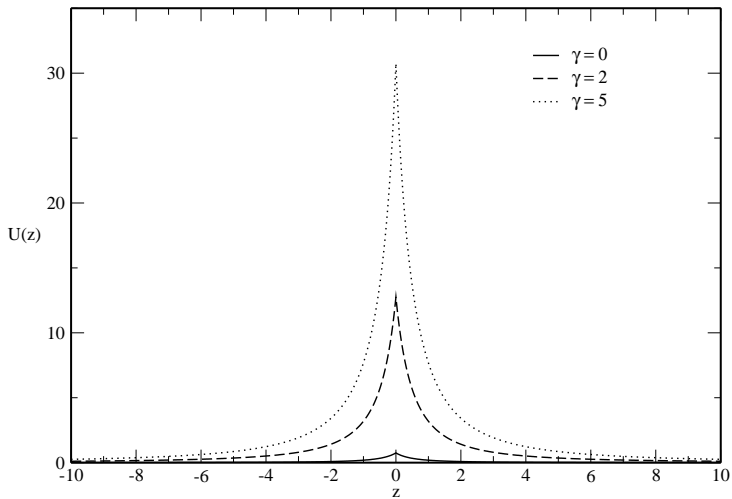}
 \caption{Plot of Schr\"odinger potential for gauge field in Randall-Sundrum scenario with $k =1$.}
 \label{fig:potGmn-RS}
\end{figure}

\begin{figure}[!h]
\centering
\subfigure[]{
\includegraphics[scale=0.2]{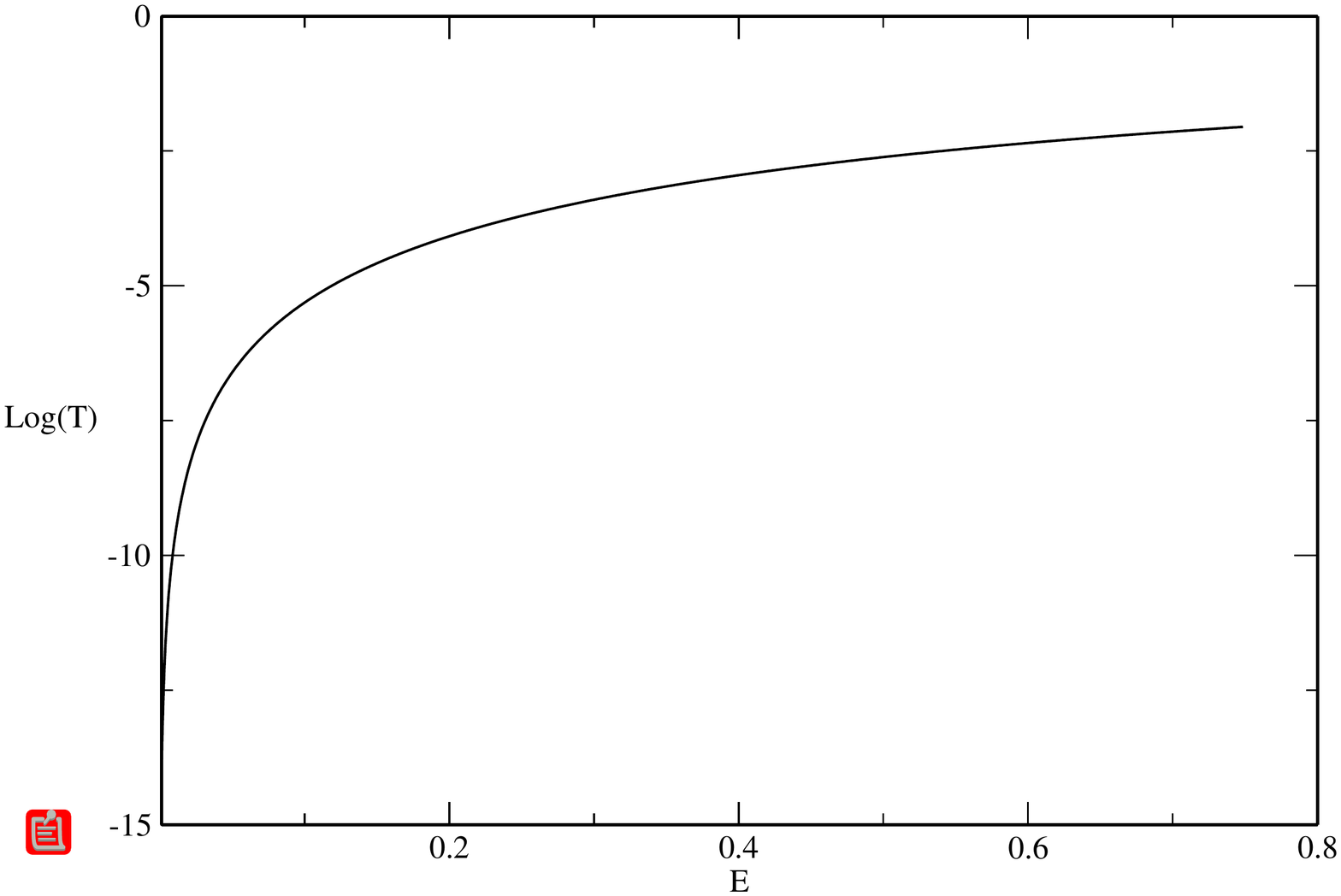}
}
\subfigure[]{
\includegraphics[scale=0.2]{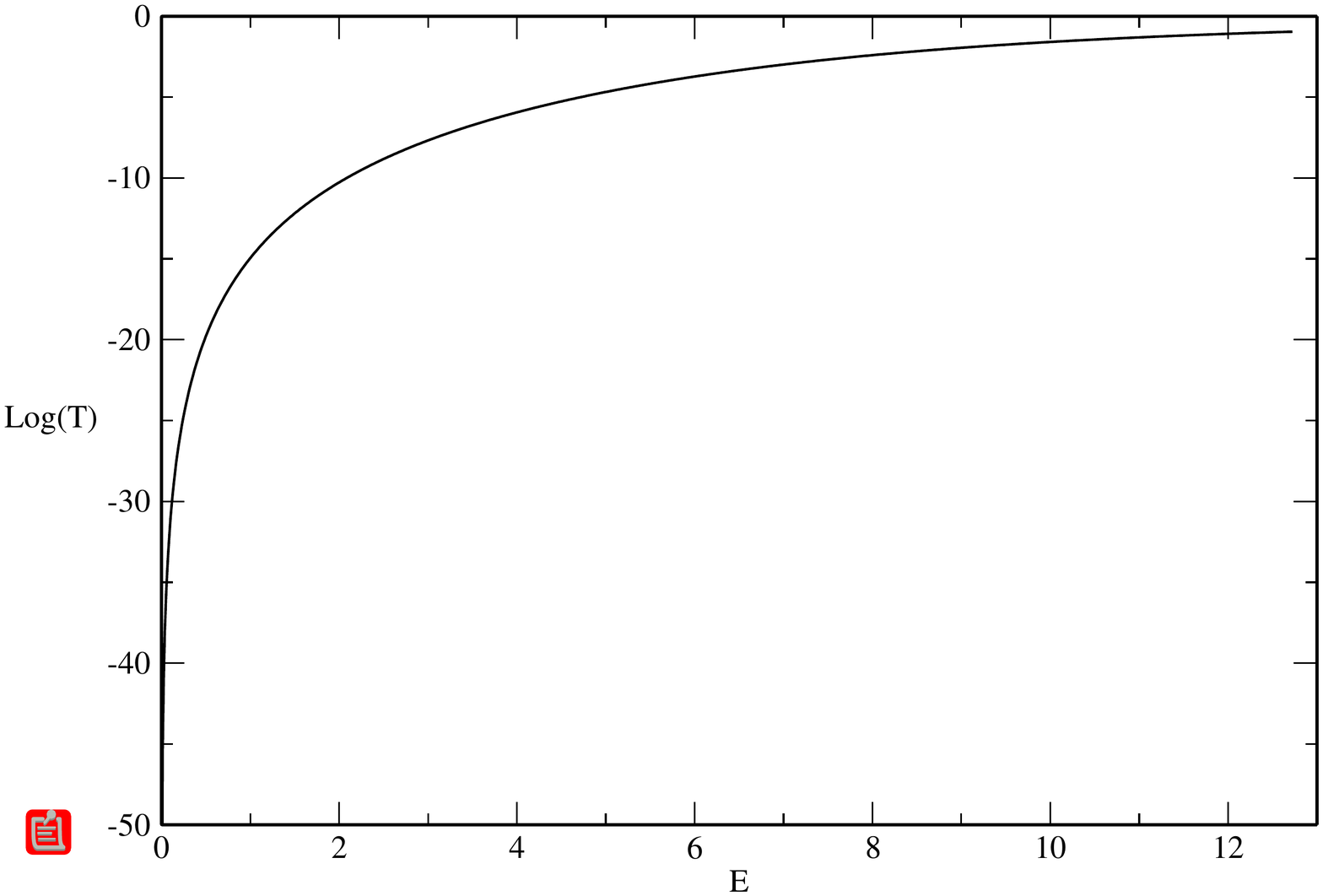}
}
\subfigure[]{
\includegraphics[scale=0.2]{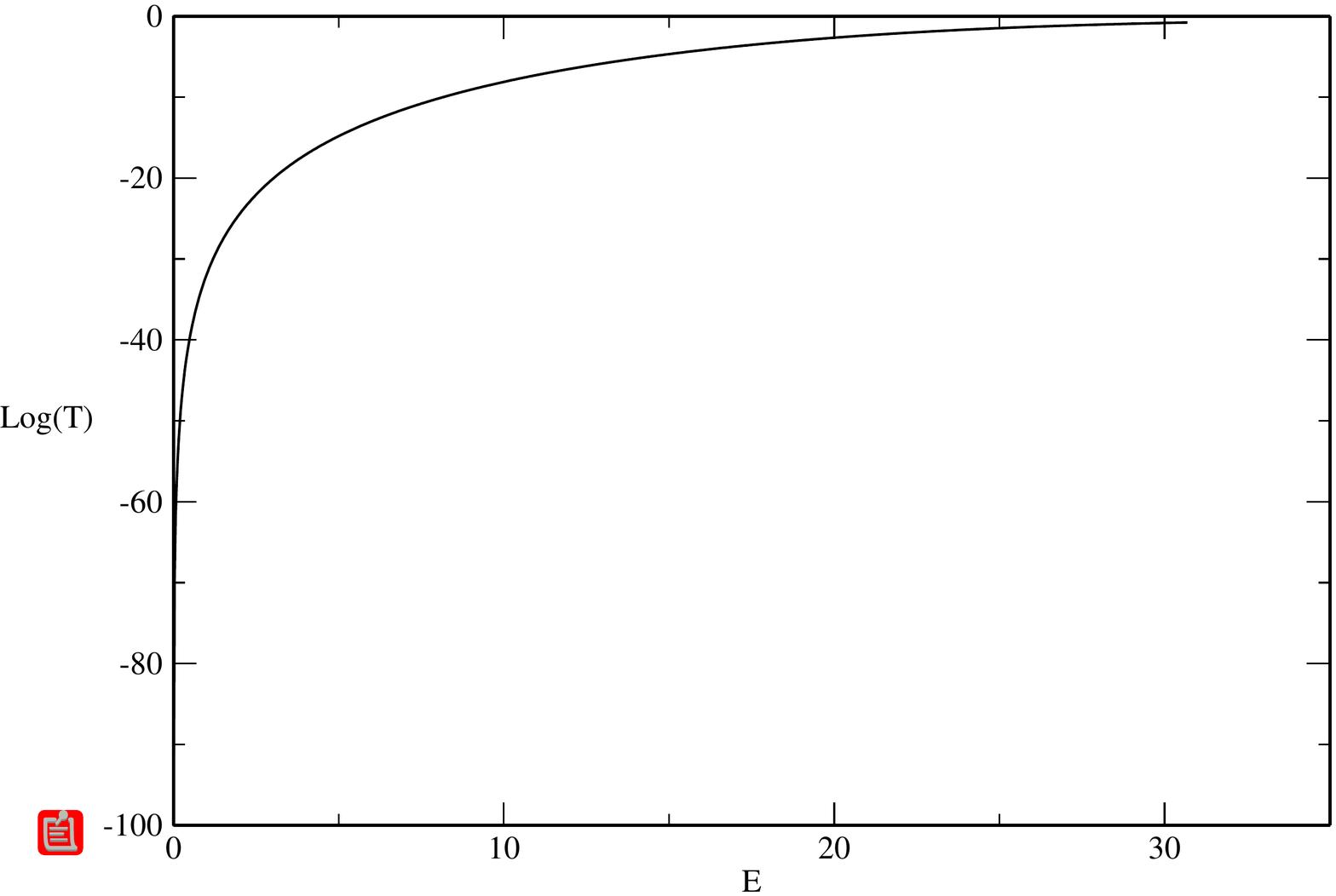}
}
\caption{Transmission coefficient for scalar field in Randall-Sundrum scenario with $k=1$ as function of energy, $E = m_{X}^{2}$. (a) For $\gamma_{2}$ = 0. (b) For $\gamma_{2}$ = 2. (c) For $\gamma_{1}$ = 5.}
\label{fig:TGmn1-RS}
\end{figure}

For the transversal part of $p$-form the potential of Schr\"odinger equation, eq. (\ref{potpG}), is given by 
\begin{equation}\label{potGRS}
U(z)=\frac{(3+24\gamma_{2})k^{2}}{4(k|z|+1)^{2}} -(1+6\gamma_{2})k\delta(z),
\end{equation}
and was illustrated in fig. \ref{fig:potGmn-RS} for some $p$-forms.
For the massive case Eq. (\ref{eqfp}) provides the solution
\begin{equation}\label{psimasGRS}
 \psi(z) =(k|z|+1)^{1/2}[C_{1}J_{\nu}(m_{X}|z|+ m_{X}/k)+C_{2}Y_{\nu}(m_{X}|z|+ m_{X}/k)],
\end{equation}
where $C_{1}$ and $C_{2}$ are constants and $\nu = \sqrt{1+6\gamma_{2}}$ . Since the Bessel functions goes to infinity as $(m_{X}|z|+ m_{X}/k)^{-1/2}$, no fixation of constants $C_{1}$ and $C_{2}$ produces a
convergent solution. In the same way of Ricci tensor case, we can write the transmission coefficient in an analytic form 
\begin{equation}
T = |\gamma|^{2} = \frac{4m_{X}^{2}}{|2F_{\nu}(0)F_{\nu}'(0) + \nu^{2} k  F_{\nu}^{2}(0)|^{2}}. 
\end{equation}
The transmission coefficient was plotted in fig. \ref{fig:TRmn-RS} as a function of energy for some $p$-forms and do not show peaks, indicating none unstable massive modes.  
\begin{figure}[!h]
\centering
\subfigure[]{
\includegraphics[scale=0.2]{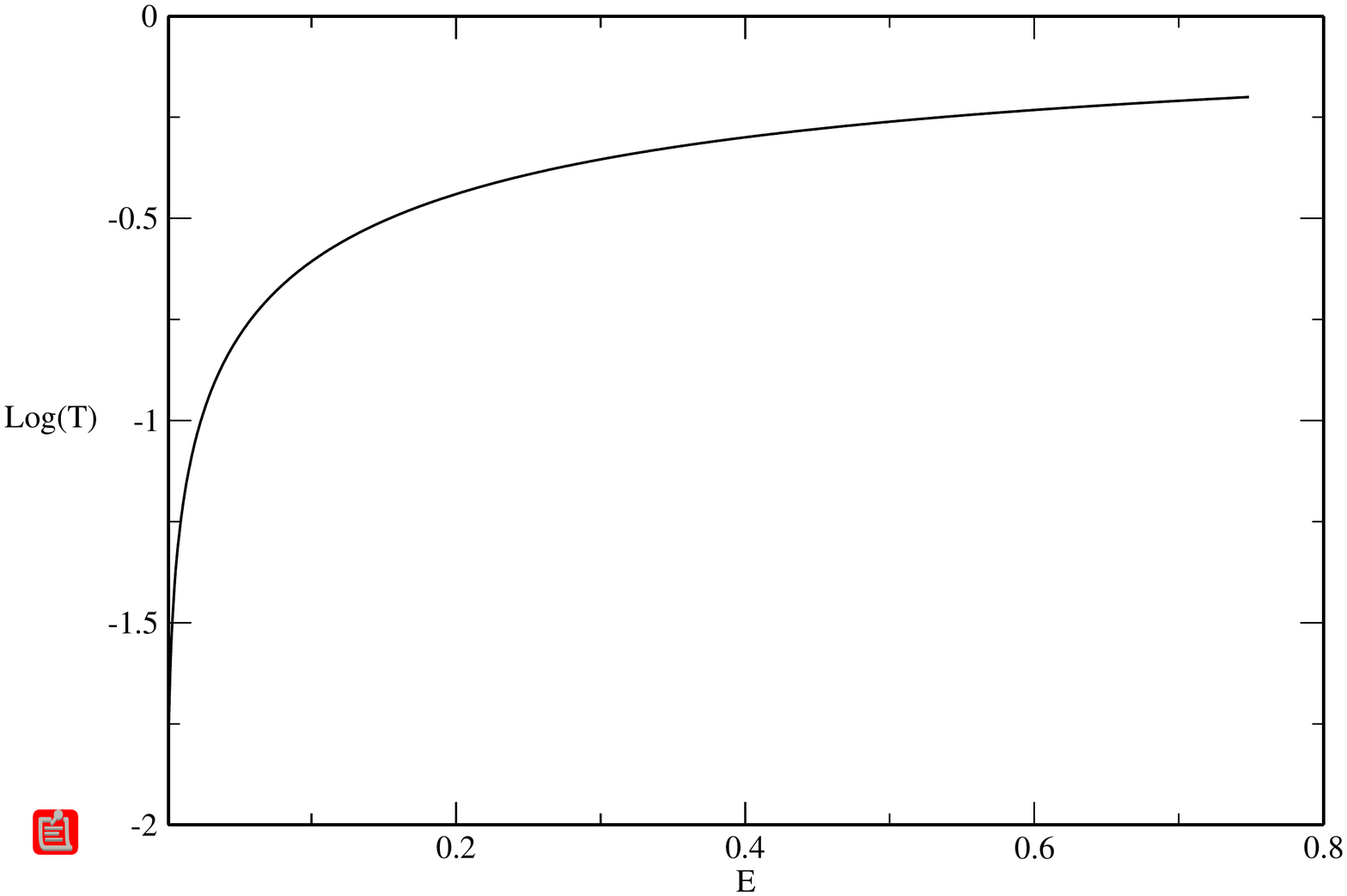}
}
\subfigure[]{
\includegraphics[scale=0.2]{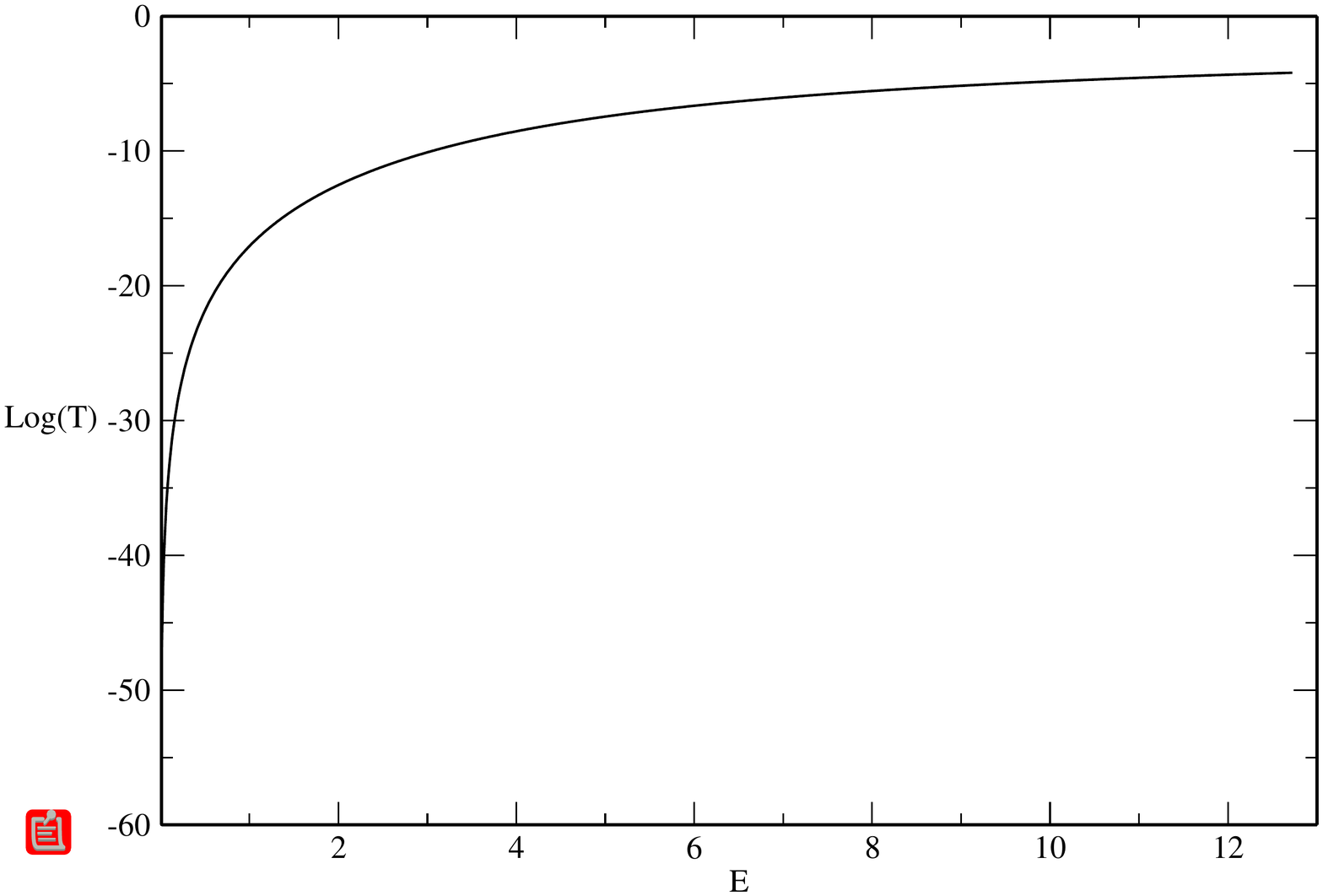}
}
\subfigure[]{
\includegraphics[scale=0.2]{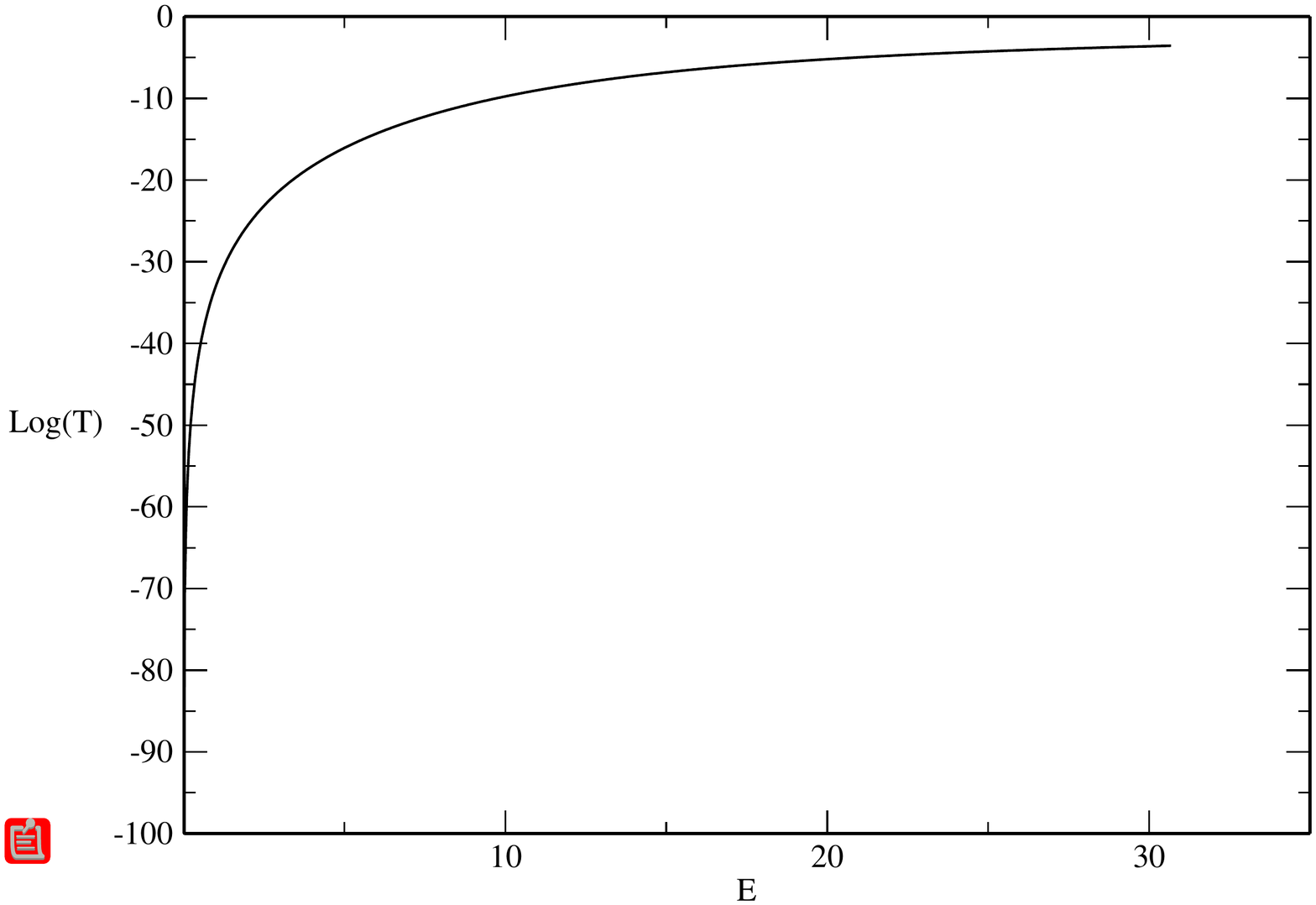}
}
\caption{Transmission coefficient for gauge field in Randall-Sundrum scenario with $k=1$ as function of energy, $E = m_{X}^{2}$. (a) For $\gamma_{2}$ = 0. (b) For $\gamma_{2}$ = 2. (c) For $\gamma_{2}$ = 5.}
\label{fig:TGmn-RS}
\end{figure}
For the scalar field  in Randall-Sundrum scenario the potential of Schr\"odinger equation, (\ref{potp-1G}), can be written as
\begin{equation}\label{potp-1GRS}
U(z)= \frac{(3 +24\gamma_{2})k^{2}}{4(k|z|+1)^{2}} -3k\delta(z).
\end{equation}
Since $G_{reg}$ in RS scenario is a constant, its does not contribute to the potential.
The potential for $(p-1)$-form is the same of $p$-form, changing only the boundary condition, the behavior of massive  modes are the same, i.e, non-localized.
The transmission coefficient was illustrated in fig. \ref{fig:TGmn1-RS}  shows the same behavior of gauge field.

\subsection{The brane scenario generated by a domain-wall}
Following the scheme used in Ricci tensor coupling, the first smooth brane scenario is the one produced by a domain-wall \cite{Du:2013bx, Melfo:2002wd}. The warp factor,
\begin{equation}
  A(z) = -\frac{1}{2n}\ln\left[\left(kz\right)^{2n}+1\right],
\end{equation}
which recover the Randall-Sundrum metric for large $z$ and $n \in N^{*}$. This brane scenario produces the following components of the Einstein tensor
\begin{eqnarray}
&&G^{\beta}_{\beta} = 12 z^{2 n-2} \left(z^{2 n}+1\right)^{\frac{1}{n}-2} \left(2 z^{2 n}-2 n+1\right) \label{Gbdw}
\\&&G^{5}_{5} = 6 z^{2 n-2} \left(z^{2 n}+1\right)^{\frac{1}{n}-2} \left(z^{2 n}-2 n+1\right). \label{G5dw}
\end{eqnarray}

\begin{figure}[!h]
\centering
\subfigure[]{
\includegraphics[scale=0.2]{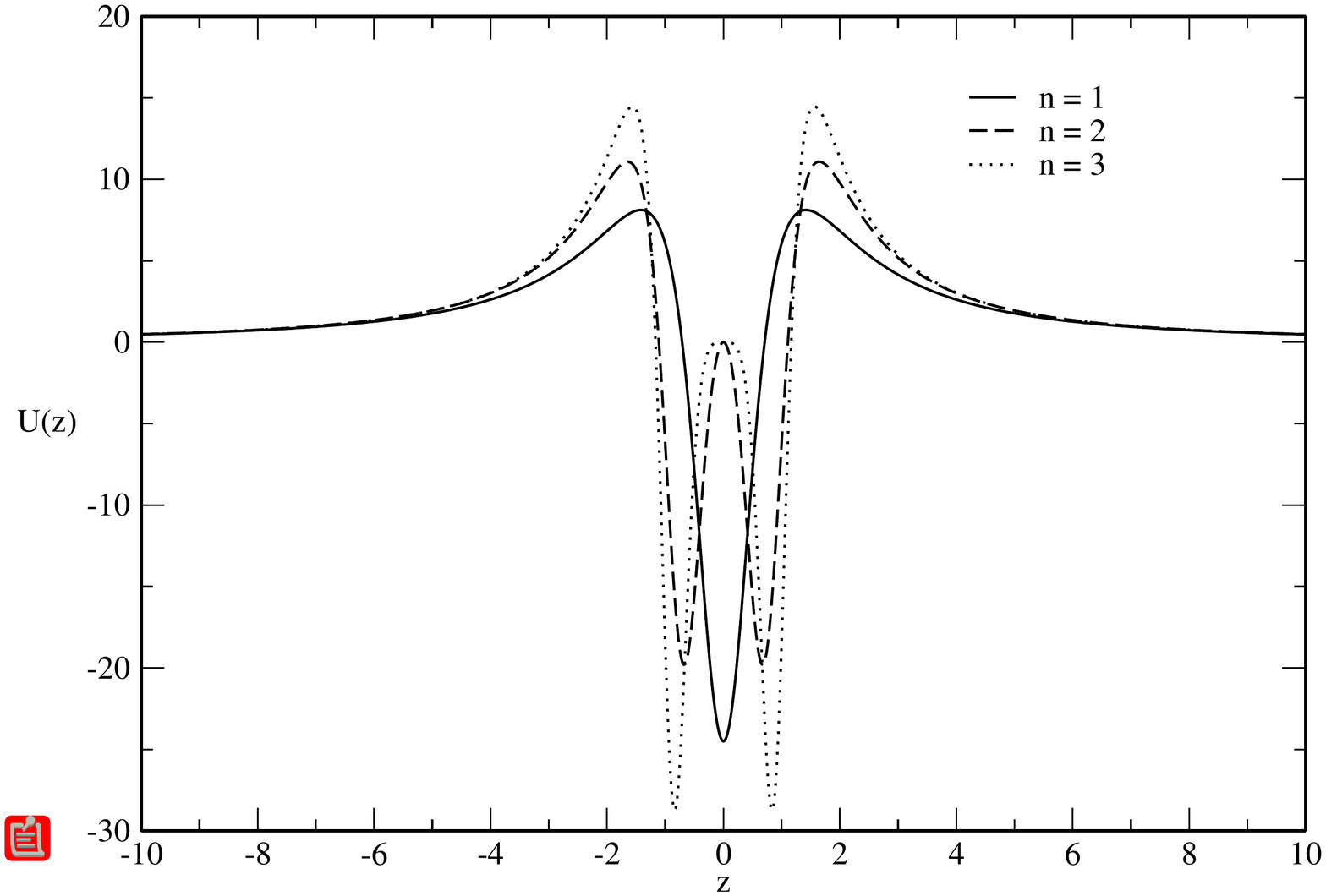}
\label{fig:PotGmn-smn}
}
\subfigure[]{
\includegraphics[scale=0.2]{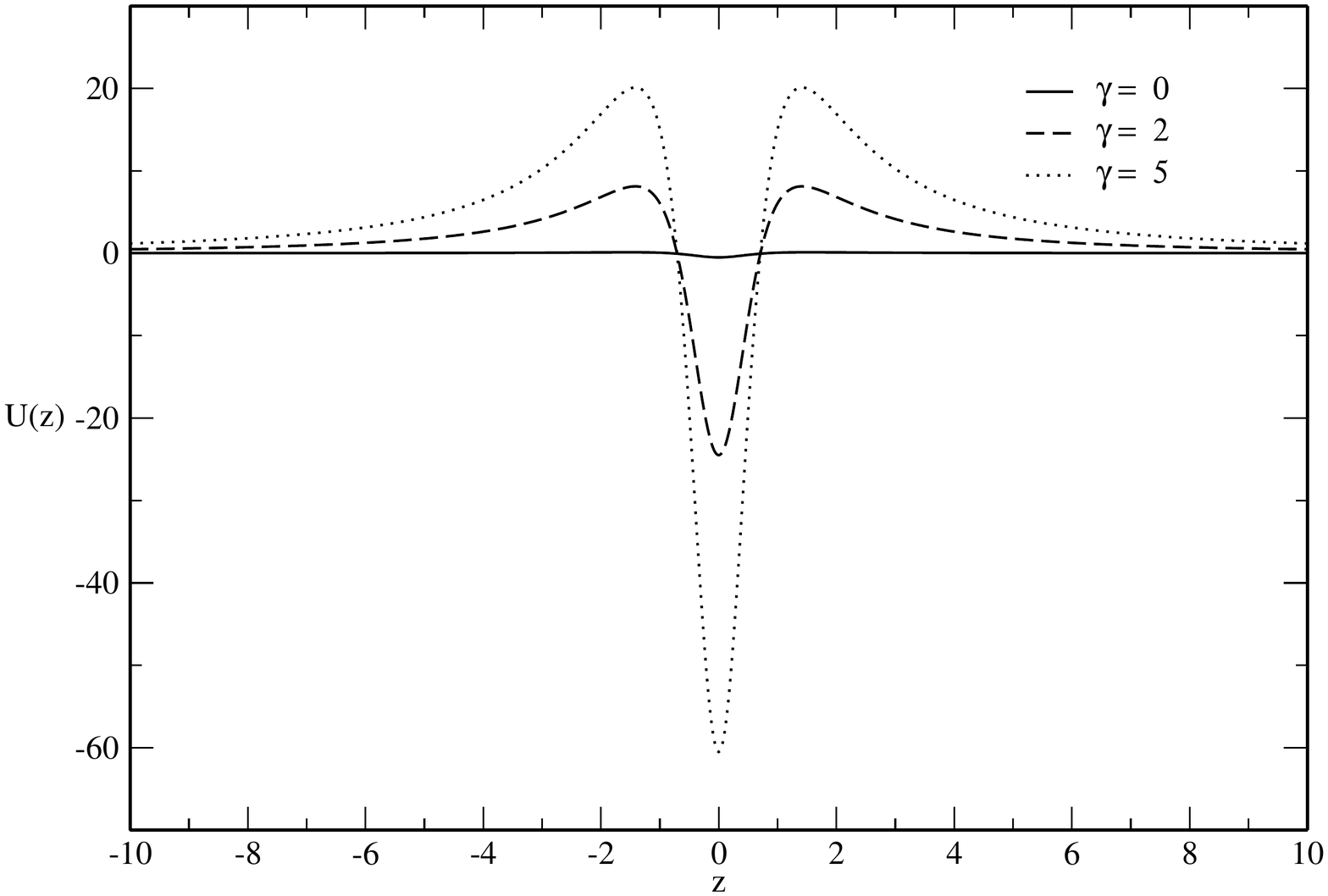}
\label{fig:PotGmn-smg}
}
\caption{ Schr\"odinger potential in smooth scenario generated by domain walls. (a) For some values of parameter $n$ with $\gamma_{2} = 2$. (b) For some values of coupling constant $\gamma_{2}$ with $n =1$.}
\end{figure}

Using this metric in eq. (\ref{potpG}) we obtain the Schr\"odinger's potential for the transversal part of $p$-form
\begin{equation}\label{potGsm}
U(z)=\frac{z^{2 n-2} \left(48 \gamma_{2} -4 (24 \gamma_{1} +1) n+(96 \gamma_{2} +3) z^{2 n}+2\right)}{4 \left(z^{2 n}+1\right)^2},
\end{equation}
which was illustrated in fig. \ref{fig:PotGmn-smn} for some values of $n$ with $\gamma_{2} = 2$ and in fig. \ref{fig:PotGmn-smg} for some values of $\gamma_{2}$ with $n =1$.

\begin{figure}[!h]
\centering
\subfigure[]{
\includegraphics[scale=0.2]{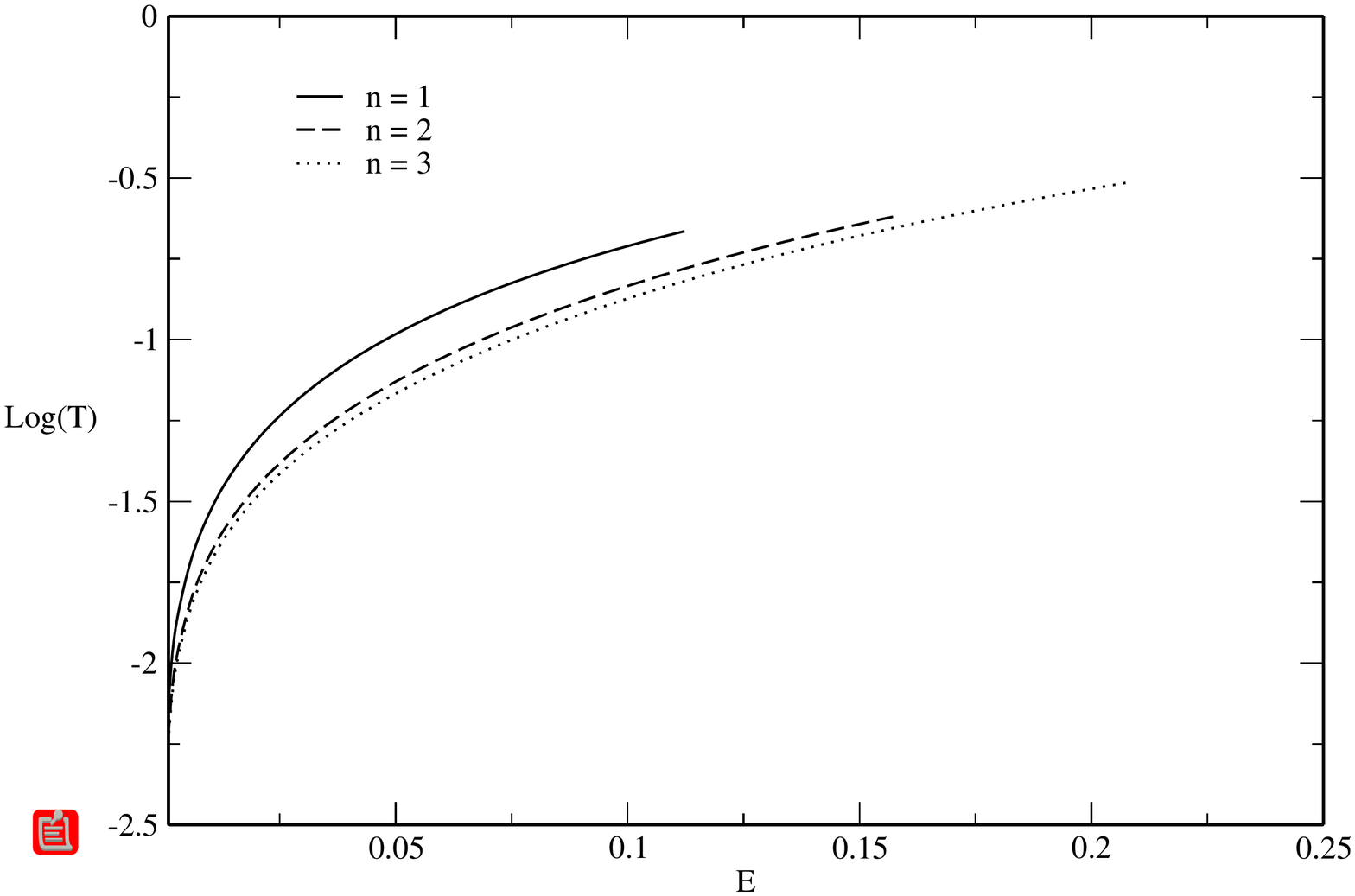}
\label{fig:TGmn-smn}
}
\subfigure[]{
\includegraphics[scale=0.2]{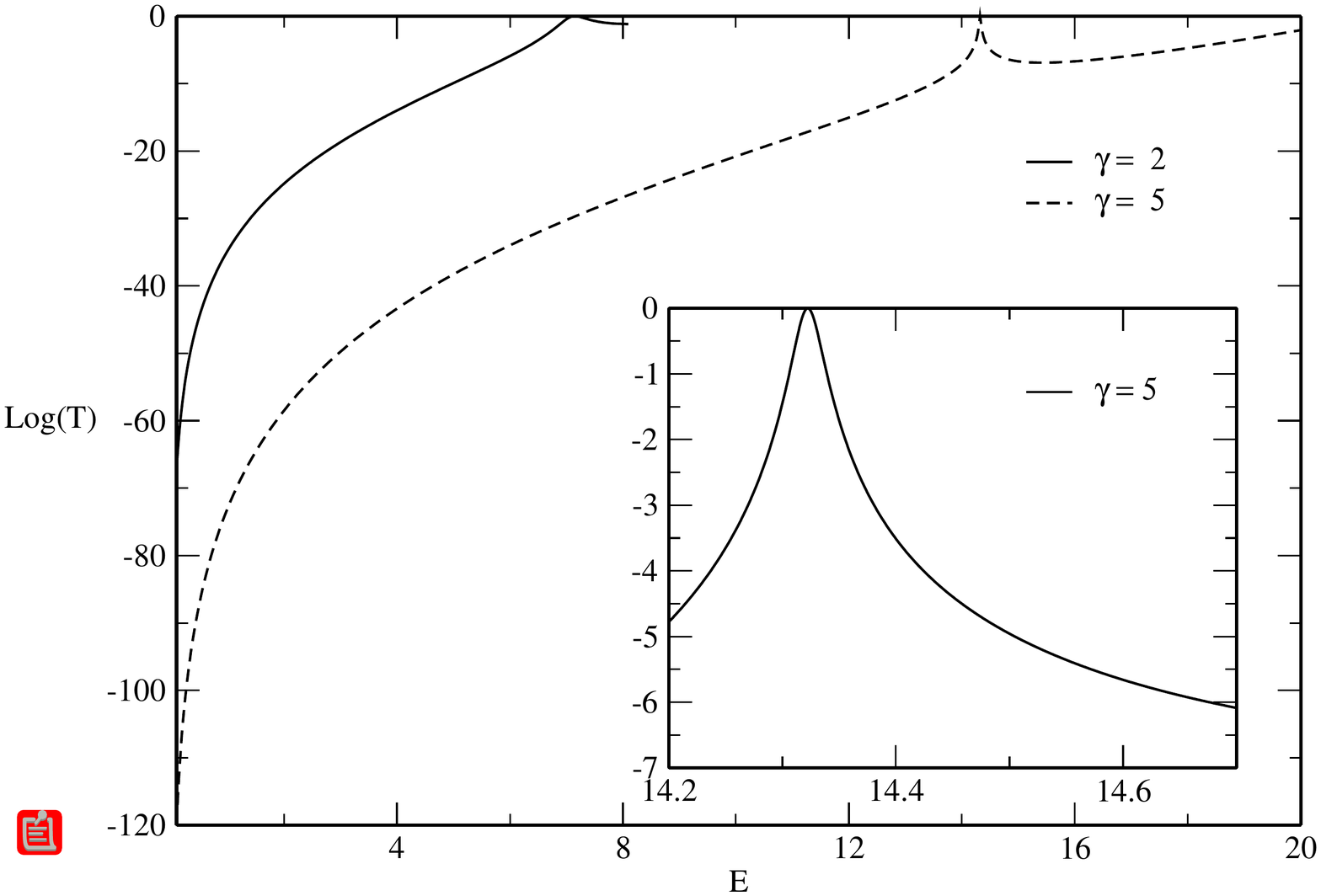}
\label{fig:TGmn-smg25}
}
\caption{Transmission coefficient for gauge field in brane scenario generated by domain-walls as function of energy, $E = m_{X}^{2}$. (a) For some values of $n$ with $\gamma_{2}$ = 0. (b) For some values of coupling constant $\gamma_{2}$ with $n=1$. As showed in detail the peak for $\gamma_{2} = 5$ is not a resonance.}
\end{figure}

The solution of massive modes for the transversal part of the gauge field can not be found analytically. To obtain information about this state we use the transfer matrix method to evaluate the transmission coefficient. 
The behavior is illustrated in fig. \ref{fig:TGmn-smn} for some values of parameter $n$ with $\gamma_{2} = 0$ and in fig. \ref{fig:TGmn-smg25} for some values of the coupling constant $\gamma_{2}$ with $n =1$. Both figures do not exhibit  peaks, indicating the absence of unstable modes.  
The inset in the figure shows that the peak for $\gamma_{2} = 5$ is not a resonance.

For massless modes of reduced scalar field the potential is given by eq. (\ref{potp-1G}). Because the components of the Einstein tensor, Eqs. (\ref{Gbdw}) and (\ref{G5dw}), vanishes at regular points near to the origin for all values of parameter
$n$, the potential diverges at these points. These kind of divergence does not allow us to use the transfer matrix method to compute the transmission coefficient and to evaluate the existence of 
unstable massive modes.
\subsection{The brane scenario generated by a kink} 
Finally we compute the transmission coefficient in another smooth brane scenario, one generated by a kink \cite{Landim:2011ki}. The warp factor is given by 
\begin{equation}
 A(y) = -4\ln\cosh y -\tanh^{2}y,
\end{equation}
where the variable $y$ relates with the conformal coordinate, $z$, by
\begin{equation}
dz = \e^ {-A(y)}dy.
\end{equation}
 The behavior of the potential of the transversal part of $p$-form field, eq. (\ref{potpG}), with the above warp factor was illustrated in fig. \ref{fig:potRmn-kink}. For massive modes, like in previous sections, we use
the transfer matrix method to compute the transmission coefficient. The results are plotted in fig. \ref{fig:TRmn-kink}, and exhibits a smooth behavior, indicating the non existence of unstable massive modes. As showed in detail
the increases of probability near to $E = 100$ is not a resonance peak.
\begin{figure}[!h]
\centering
\subfigure[]{
\includegraphics[scale=0.2]{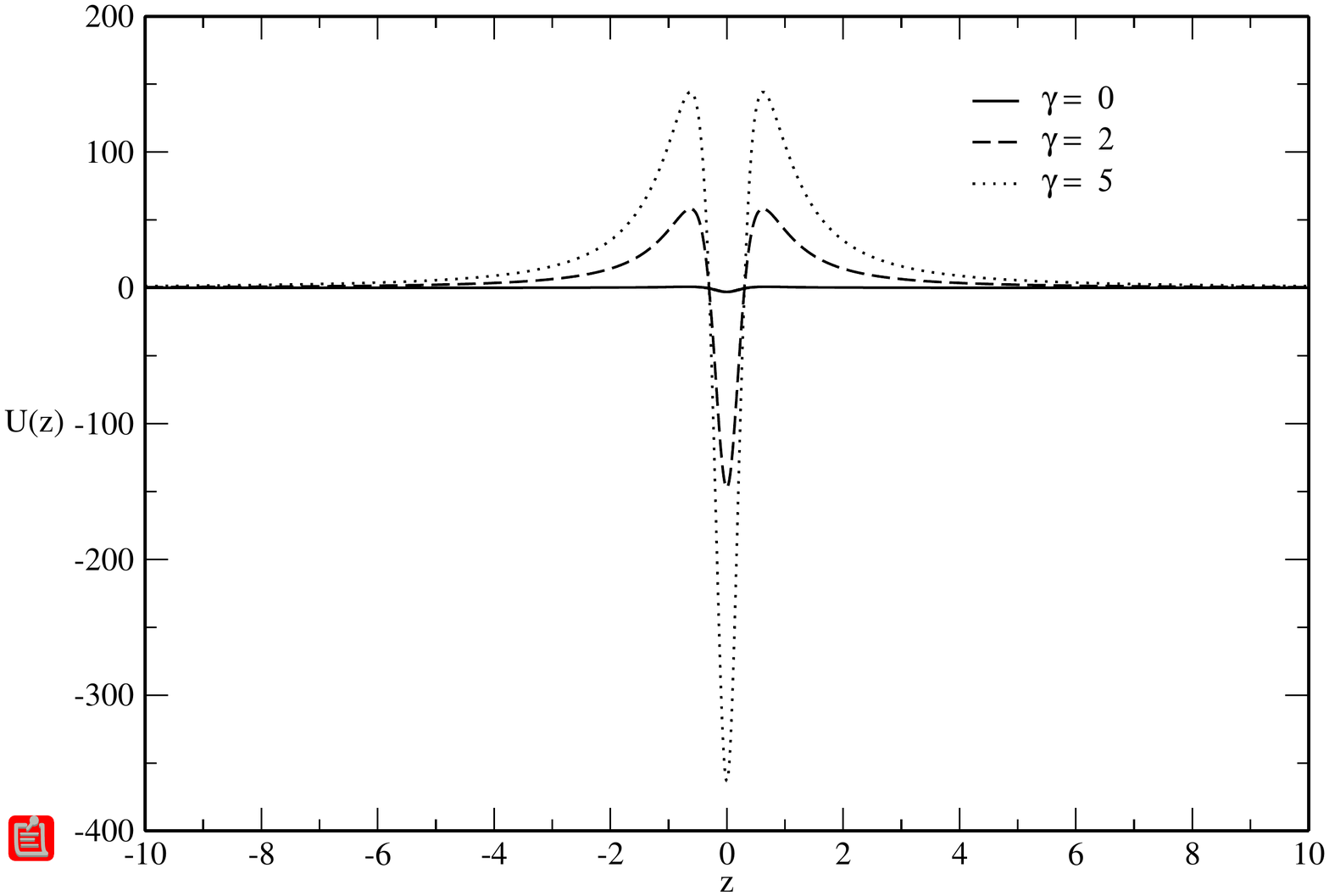}
\label{fig:potGmn-kink}
}
\subfigure[]{
\includegraphics[scale=0.2]{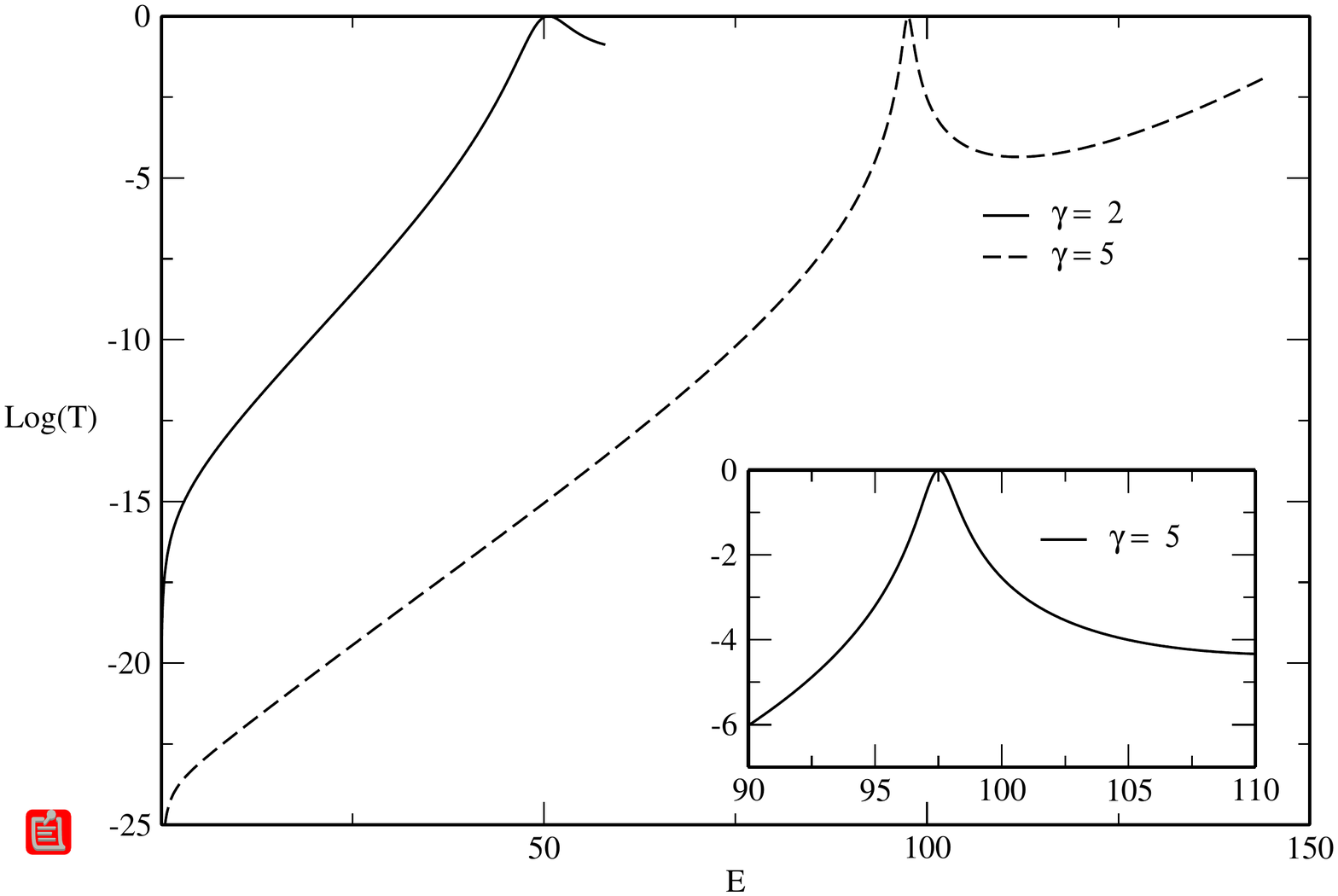}
\label{fig:TGmn-kinkg}
}
\caption{(a) Schr\"odinger potential in kink scenario for gauge field for some values of coupling constant $\gamma_{2}$. (b) Transmission coefficient in kink scenario for gauge field as function of energy, $E= m_{X}^{2}$. As showed in detail the peak for $\gamma_{2} = 5$ is not a resonance.}
\end{figure}
For the scalar field, like the previous case, the components of Einstein tensor vanishes at regular points, making the potential divergent at this same points. These kind of divergence does not allow us to use the transfer matrix method to compute the transmission coefficient and to evaluate the existence of 
unstable massive modes.

\section{Conclusions}
In this paper we have generalized the geometrical localization mechanism. The strongest point of the model is that it solves the problem of gauge field localization without the addition of new degrees of freedom. The generalization was made by considering other kinds of couplings to the mass term of a vector field.  Since the mass term is quadratic and symmetric, we must consider only tensors with two symmetric indices. The possibilities are the Ricci and the Einstein tensors. Since the mass term breaks gauge invariance
we must show that it is recovered for the zero mode. For this we first split the vector field in its transversal and longitudinal components. For both couplings
we then show that the components are decoupled in the equations of motion. This provides a consistent four dimensional, gauge invariant, equation of motion for the zero mode of the vector field. After this we analyzed the localizability of this zero mode. This is important for both RS models, providing a consistent way of obtaining a four dimensional, gauge invariant action, from a bulk action.   We discovered that the Ricci tensor provides us with a localized zero mode if we fix the coupling constant to $\gamma_{2}=-2$. Since the result is valid for any range in the integration and for any warp factor it is valid for RS I and II. For the Einstein tensor we showed that the solution is not localized. 

Another important point of the paper is the discussion about the kind
of couplings that can generate localization.  The fact that $g^{MN}$ and $G^{MN}$ do not generate localized zero modes is at least curious.
Both have a null divergence. However, $g^{MN}R$ and $R^{MN}$ generate
a localized zero mode and these tensors do not have a null divergence. 
The above results points to the fact that tensors with null divergence can 
not generate localized zero modes for the gauge field. Perhaps this is pointing to some yet unseen fundamental property of this kind of coupling. Until a prove of this is found in general, it must be tested case by case.   

Despite the fact that the zero modes are localized, unstable massive modes can be seen over the membrane. For this analysis we have computed the transmission coefficient using the transfer matrix in order to find resonance peaks indicating the existence of these modes. We considered three different brane scenarios: Delta-like, Kink-like and Domain-wall-like. The first case considered was the coupling with the Ricci tensor. In RS, or delta-like, brane scenario the analytical solution of the Schr\"odinger equation for massive modes shows that the transmission coefficient does not exhibits resonance peaks.  Also in this brane model we showed that the scalar field has no unstable massive modes on the brane, i.e, the transmission coefficient does not exhibits resonance peaks.

Next the smooth scenarios were considered. In booth scenarios the Schr\"odinger equation can not be solved analytically and we used the transfer matrix method to compute the transmission coefficient. The results indicated that no unstable massive modes of vector field can be found on the brane. Also in smooth branes the computational method could no be applied to compute the transmission coefficient of scalar field. This is due to the fact that the Ricci tensor vanish at regular points in booth smooth scenarios, making the potential divergent at these points.

Finally, the massive modes for the coupling with the Einstein tensor were considered. We have shown that this coupling does not localize the massive and the zero modes of vector field. Anyway we searched for unstable massive modes by computing the transmission coefficient. Since the zero mode is non-localized, the coupling constant is not fixed, like in the Ricci tensor case. This freedom allowed us to compute the transmission coefficient for some values of the coupling constant. Like in previous cases, in RS scenario the transmission coefficient could be computed analytically for the vector field and for scalar field. Booth coefficients does not exhibits resonances peaks, i.e., no unstable modes of these fields can be found on the brane for all parameters used. As before, we considered the smooth brane cases. Despite the presence of a most probable states for $\gamma =5$, the result indicated that no resonance peaks appears for all parameters used in the numeric calculation. Like in Ricci tensor coupling case,
the potential for scalar field diverges at regular points, making impossible the use of the transfer matrix method to compute the transmission coefficient in booth smooth cases. With all the above calculations got to the conclusion that geometrical couplings do not generate resonances. In a previous study some of us has considered the resonances for the Ricci scalar case where this same pattern has been found. We should point that, just as in the analysis of the zero mode, until some general proof of this pattern is found, it must be tested case by case.  
\section*{Acknowledgments}

We acknowledge the financial support provided by Funda\c c\~ao Cearense de Apoio ao Desenvolvimento Cient\'\i fico e Tecnol\'ogico (FUNCAP), the Conselho Nacional de 
Desenvolvimento Cient\'\i fico e Tecnol\'ogico (CNPq) and FUNCAP/CNPq/PRONEX.

\end{document}